\newcommand\T{\rule{0pt}{2.6ex}}
\newcommand\B{\rule[-1.2ex]{0pt}{0pt}}
\documentclass[PAPER, cernpreprint, atlasdraft=false, texlive=2023, UKenglish, texmf, orcidlogo]{atlasdoc} 
 
\usepackage{atlaspackage}
\usepackage{float}
\usepackage{multirow}
\usepackage{lscape}
\usepackage{verbatim}
\usepackage{atlasbiblatex}
 
\usepackage{atlasphysics}
 
\graphicspath{{logos/}{figures/}}

\AtlasTitle{Determination of the strong coupling constant from transverse energy--energy correlations in multijet events at $\sqrt{s} = 13$ TeV with the ATLAS detector}
 
\AtlasAbstract{
Measurements of transverse energy--energy correlations and their associated azimuthal asymmetries in multijet events are presented. The analysis is performed using a data sample corresponding to 139~$\mbox{fb\(^{-1}\)}$ of proton--proton collisions at a centre-of-mass energy of $\sqrt{s} = 13$ TeV, collected with the ATLAS detector at the Large Hadron Collider. The measurements are presented in bins of the scalar sum of the transverse momenta of the two leading jets and unfolded to particle level. They are then compared to next-to-next-to-leading-order perturbative QCD calculations for the first time, which feature a significant reduction in the theoretical uncertainties estimated using variations of the renormalisation and factorisation scales. The agreement between data and theory is good, thus providing a precision test of QCD at large momentum transfers $Q$. The strong coupling constant $\alpha_s$ is extracted as a function of $Q$, showing a good agreement with the renormalisation group equation and with previous analyses. A simultaneous fit to all transverse energy--energy correlation distributions across different kinematic regions yields a value of $\alpha_\mathrm{s}(m_\ensuremath{Z}) = 0.1175 \pm 0.0006 \mbox{ (exp.)} ^{+0.0034}_{-0.0017} \mbox{ (theo.)}$, while the global fit to the asymmetry distributions yields $\alpha_{\mathrm{s}}(m_\ensuremath{Z}) = 0.1185 \pm 0.0009 \mbox{ (exp.)} ^{+0.0025}_{-0.0012} \mbox{ (theo.)}$.
}
 
\author{The ATLAS Collaboration}
 
\AtlasRefCode{STDM-2018-51}
\AtlasJournalRef{JHEP 07 (2023) 85}
\AtlasDOI{10.1007/JHEP07(2023)085}
\PreprintIdNumber{CERN-EP-2022-282}

\hypersetup{pdftitle={ATLAS document},pdfauthor={The ATLAS Collaboration}}
 
\begin{document}
\maketitle

\section{Introduction}
Multijet final states, produced in proton--proton ($pp$) collisions with large momentum transfer at the LHC provide an ideal testing ground for perturbative Quantum Chromodynamics (pQCD). Event shapes~\cite{evShapes1, evShapes2} are a class of observables defined as functions of the final-state particles four-momenta, which characterise the hadronic energy flow in a collision. They can be used to precisely test pQCD calculations and additionally to extract the value of the strong coupling constant, $\alpha_{\mathrm{s}}$. Event shape variables were measured in $e^+ e^-$ collision experiments from PETRA and PEP~\cite{plutoEventShapes,tassoEventShapes,jadeEventShapes} to LEP and SLC~\cite{opalEventShapes,alephEventShapes,thrustDelphi,l3eventShapes,SLD}, at the $ep$ collider HERA~\cite{zeusEventShapes1,zeusEventShapes2,h1EventShapes1,h1EventShapes2,h1EventShapes3} and in hadron--hadron collisions at the Tevatron~\cite{Aaltonen} and the LHC~\cite{Khachatryan, Aad, evShapes}.
 
A particularly interesting, infrared and collinear safe, event-shape observable is the energy--energy correlation (EEC) function, which was originally introduced for $e^{+} e^{-}$ annihilation experiments~\cite{EEC-TH-00, EEC-TH-01}. The EEC function and its associated azimuthal asymmetry (AEEC) can be calculated in pQCD, and the $\mathcal{O}(\alpha^{2}_{\mathrm{s}})$ corrections are modest~\cite{EEC-TH-02, EEC-TH-03, EEC-TH-04, EEC-TH-04b, DESY-88-081}. They were also studied in the nearly back-to-back limit~\cite{EEC-TH-05}. The EEC measurements~\cite{PLUTO, MARKII, MARKJ, CELLO, JADE, MAC, TASSO, TOPAZ, DELPHI, OPAL, ALEPH, L3, eecSLD} have had significant impact on the early precision tests of QCD and in the determination of the strong coupling constant.
 
Transverse energy--energy correlations (TEEC) and their associated azimuthal asymmetries (ATEEC) were proposed as the appropriate generalisation for hadron collider experiments in Ref.~\cite{CERN-TH-3800}, where leading-order (LO) predictions were also presented. In experiments with incoming hadrons, observables that are invariant with respect to boosts along the direction of the beam axis are beneficial. As jet-based observables, the TEEC and ATEEC make use of the jet transverse energy $E_{\mathrm{T}} = E / \cosh y$, where $E$ is the jet energy and $y$ is the jet rapidity. The TEEC function is defined as the transverse-energy-weighted distribution of the azimuthal differences between jet pairs in the final state~\cite{TEEC-TH-00}, i.e.
\begin{equation}
\frac{1}{\sigma}\frac{\mathrm{d}\Sigma}{\mathrm{d}\cos\phi} \equiv \frac{1}{\sigma}\sum_{ij}\int\frac{\mathrm{d}\sigma}{\mathrm{d}x_{\mathrm{T}i} \mathrm{d}x_{\mathrm{T}j} \mathrm{d}\cos\phi}x_{\mathrm{T}i}x_{\mathrm{T}j} \mathrm{d}x_{\mathrm{T}i} \mathrm{d}x_{\mathrm{T}j} = \frac{1}{N} \sum_{A=1}^N \sum_{ij} \frac{E^A_{\mathrm{T}i}E^A_{\mathrm{T}j}}{\left(\sum_k E^A_{\mathrm{T}k}\right)^2}\delta(\cos \phi - \cos\varphi_{ij}),
\label{eq:teecDef}
\end{equation}
where the last expression is valid for a sample of $N$ multijet events, labelled by the index $A$, and the indices $i$, $j$ and $k$ run over all jets in a given event. Here, $x_{\mathrm{T}i} = E_{\mathrm{T}i} / \sum_k E_{\mathrm{T}k}$ is the normalised transverse energy of jet $i$, $\varphi_{ij}$ is the angle in the transverse plane between jet $i$ and jet $j$ and $\delta(x)$ is the Dirac delta function, which ensures $\phi = \varphi_{ij}$. The distribution is normalised to the total cross-section, $\sigma$.
 
To cancel out uncertainties that are symmetric in $\cos\phi$, the ATEEC function is defined as the difference between the forward ($\cos\phi > 0$) and the backward ($\cos\phi < 0$) part of the TEEC function, i.e.
\[
\frac{1}{\sigma} \frac{\mathrm{d}\Sigma^{\mathrm{asym}}}{\mathrm{d}\cos\phi}  =
\frac{1}{\sigma} \frac{\mathrm{d}\Sigma}{\mathrm{d}\cos\phi} \biggr\rvert_\phi -
\frac{1}{\sigma} \frac{\mathrm{d}\Sigma}{\mathrm{d}\cos\phi} \biggr\rvert_{\pi-\phi} .
\]
Both the TEEC and ATEEC functions are sensitive to gluon radiation and show a clear dependence on the strong coupling. Numerical results at next-to-leading order (NLO) for the jet-based TEEC function were obtained~\cite{TEEC-TH-01} by using the \textsc{Nlojet++} program~\cite{IPPP/01/48, TEEC-TH-03}, which provides the LO and NLO calculations for three-jet production. Furthermore, numerical results for the hadron-based TEEC function at next-to-leading order plus next-to-next-to-leading-logarithmic (NLO+NNLL) accuracy were computed recently~\cite{TEEC-TH-02}.
 
The ATLAS Collaboration has presented measurements of the TEEC and ATEEC functions at centre-of-mass energies of $\sqrt{s} = 7~\TeV$~\cite{STDM-2014-03} and $\sqrt{s} = 8~\TeV$~\cite{STDM-2016-10}, determining $\alpha_{\mathrm{s}}$ as a function of the interaction scale and testing the running of $\alpha_{\mathrm{s}}$ predicted by the renormalisation group equation (RGE). The existence of new coloured fermions would imply modifications to the QCD $\beta$-function~\cite{TEEC-TH-04, TEEC-TH-05}. Therefore, the running of $\alpha_{\mathrm{s}}$ is not only important as a precision test of QCD at large scales but also as a test for new physics. An interpretation of these measurements in terms of constraints on new coloured particles through their impact on the running of $\alpha_{\mathrm{s}}$ is presented in Ref.~\cite{TEEC-TH-06}.
 
This analysis presents a measurement of the TEEC and ATEEC functions at a centre-of-mass energy of $\sqrt{s} = 13~\TeV$ using 139~\ifb of $pp$ collision data, recorded by the ATLAS detector. This extends the energy range of the measurement with respect to previous analyses and improves the experimental precision on these observables.
 
The publication of the next-to-next-to-leading-order (NNLO) corrections to three-jet production in $pp$ collisions~\cite{mitov2022,mitov2023} provides a significant theoretical improvement in the precision of these observables. In particular, the theoretical uncertainties are significantly reduced with the addition of higher-order corrections. This allows more precise tests of pQCD and an important reduction of the uncertainties in the determination of the strong coupling constant $\alpha_s$.


\section{ATLAS detector}
\newcommand{\AtlasCoordFootnote}{ATLAS uses a right-handed coordinate system with its origin at the nominal interaction point (IP) in the centre of the detector and the \(z\)-axis along the beam pipe. The \(x\)-axis points from the IP to the centre of the LHC ring, and the \(y\)-axis points upwards. Cylindrical coordinates \((r,\phi)\) are used in the transverse plane, \(\phi\) being the azimuthal angle around the \(z\)-axis. The pseudorapidity is defined in terms of the polar angle \(\theta\) as \(\eta = -\ln \tan(\theta/2)\).}
 
The ATLAS detector~\cite{PERF-2007-01} at the LHC covers nearly the entire solid angle around the collision point.\footnote{\AtlasCoordFootnote}
It consists of an inner tracking detector surrounded by a thin superconducting solenoid, electromagnetic and hadron calorimeters,
and a muon spectrometer incorporating three large superconducting air-core toroidal magnets.
 
The inner-detector system (ID) is immersed in a \SI{2}{\tesla} axial magnetic field
and provides charged-particle tracking in the range of \(|\eta| < 2.5\).
The high-granularity silicon pixel detector covers the vertex region and typically provides four measurements per track,
the first hit normally being in the insertable B-layer (IBL) installed before Run~2~\cite{ATLAS-TDR-19,PIX-2018-001}.
It is followed by the silicon microstrip tracker (SCT), which usually provides eight measurements per track.
These silicon detectors are complemented by the transition radiation tracker (TRT),
which enables radially extended track reconstruction up to \(|\eta| = 2.0\).
The TRT also provides electron identification information
based on the fraction of hits (typically 30 in total) above a higher energy-deposit threshold corresponding to transition radiation.
 
The calorimeter system covers the pseudorapidity range of \(|\eta| < 4.9\).
Within the region \(|\eta|< 3.2\), electromagnetic calorimetry is provided by barrel and
endcap high-granularity lead/liquid-argon (LAr) calorimeters,
with an additional thin LAr presampler covering \(|\eta| < 1.8\)
to correct for energy loss in material upstream of the calorimeters.
Hadron calorimetry is provided by the steel/scintillator-tile calorimeter,
segmented into three barrel structures with \(|\eta| < 1.7\), and two copper/LAr hadron endcap calorimeters.
The solid-angle coverage is completed with forward copper/LAr and tungsten/LAr calorimeter modules
optimised for electromagnetic and hadronic energy measurements respectively.
 
The muon spectrometer (MS) comprises separate trigger and
high-precision tracking chambers measuring the deflection of muons in a magnetic field generated by the superconducting air-core toroidal magnets.
The field integral of the toroids ranges between \num{2.0} and \SI{6.0}{\tesla\metre}
across most of the detector.
Three layers of precision chambers, each consisting of layers of monitored drift tubes, covers the region \(|\eta| < 2.7\),
complemented by cathode-strip chambers in the forward region, where the background is highest.
The muon trigger system covers the range of \(|\eta| < 2.4\) with resistive-plate chambers in the barrel, and thin-gap chambers in the endcap regions.
 
Interesting events are selected by the first-level trigger system implemented in custom hardware,
followed by selections made by algorithms implemented in software in the high-level trigger~\cite{TRIG-2016-01}.
The first-level trigger accepts events from the \SI{40}{\MHz} bunch crossings at a rate below \SI{100}{\kHz},
which the high-level trigger further reduces to record events to disk at about \SI{1}{\kHz}. An extensive software suite~\cite{ATL-SOFT-PUB-2021-001} is used in data simulation, in the reconstruction and analysis of real and simulated data, in detector operations, and in the trigger and data acquisition
systems of the experiment.


\section{Data and Monte Carlo samples}
\label{sec:MC_simulation}
The data used in this analysis was collected from 2015 to 2018 at a centre-of-mass energy of $\sqrt{s} = 13~\TeV$. After applying quality criteria to ensure good ATLAS detector operation, the total integrated luminosity useful for data analysis is 139~$\ifb$~\cite{DAPR-2018-01}. The average number of inelastic $pp$ interactions produced per bunch crossing, hereafter referred to as `pile-up', is $\langle\mu\rangle = 33.6$.
 
Several Monte Carlo (MC) simulated samples are used for this analysis; they differ in the matrix element calculation (ME) and/or the parton shower (PS). The samples were produced using the \textsc{Pythia 8} \cite{Sjostrand:2006za, Sjostrand:2014zea}, \textsc{Sherpa} \cite{sherpa,sherpa22} and \textsc{Herwig 7} \cite{Bahr:2008pv,Bellm:2015jjp,Bellm:2017bvx} generators. The main features of the samples are summarised in Table~\ref{tab:mcSamples}.
\begin{table}[H]
\small
\caption{Properties of the MC samples used in the analysis, including the perturbative order in $\alpha_{\mathrm{s}}$, the number of final-state partons, the PDF set, the parton shower algorithm, the renormalisation and factorisation scales $\mu_R$ and $\mu_F$, and the value of $\alpha_{\mathrm{s}}(m_\Zboson)$ for the matrix element. \textsc{Herwig 7} was used for the generation of two samples, differing by their treatment of the parton shower.}
\label{tab:mcSamples}
\begin{center}
\begin{tabular}{c|c|c|c|c|c|c}
\T\B Generator & ME order & ME partons & PDF set & Parton shower & Scales $\mu_R,\mu_F$ & $\alpha_{\mathrm{s}}(m_\Zboson)$\\
\hline
\hline
\T\T\B \textsc{Pythia 8} & LO & 2 & \textsc{NNPDF 2.3 LO} & $\pt$-ordered & $(m^2_{\mathrm{T}3}\cdot m^2_{\mathrm{T}4})^\frac{1}{2}$ & 0.140\\
\hline
\multirow{2}{*}{\T\B \textsc{Sherpa}} & \multirow{2}{*}{LO} & \multirow{2}{*}{2,3} & \multirow{2}{*}{\textsc{CT14 NNLO}} & \multirow{2}{*}{CSS (dipole)} & \T\B $H(s,t,u)$ [$2\to 2$] & \multirow{2}{*}{0.118}\\
&    &     &                    &              & CMW [$2\to 3$]                       & \\
\hline
\multirow{2}{*}{\T\B \textsc{Herwig 7}} & \multirow{2}{*}{NLO} & \multirow{2}{*}{2,3} & \multirow{2}{*}{\textsc{MMHT2014 NLO}} & \T\B Angular-ordered & \multirow{2}{*}{$\max\left\{p_{\mathrm{T}i}\right\}_{i=1}^N$} & \multirow{2}{*}{0.120}\\
&                                    &  & & Dipole &\\
\end{tabular}
\end{center}
\end{table}
The \textsc{Pythia 8} sample was generated using \textsc{Pythia 8.235}. The matrix element was calculated for the $2\to 2$ process. The parton shower algorithm included initial- and final-state radiation based on the dipole-style $\pt$-ordered evolution, including $\gamma\to q\bar{q}$ branchings and a detailed treatment of the colour connections between partons~\cite{Sjostrand:2006za}. The renormalisation and factorisation scales were set to the geometric mean of the squared transverse masses of the two outgoing particles (labelled 3 and 4), i.e. $\sqrt{m^2_{\mathrm{T}3} \cdot m^2_{\mathrm{T}4}}$ = $\sqrt{(\pt^2 + m_3^2) \cdot (\pt^2 + m_4^2)}$. The \textsc{NNPDF 2.3 LO} parton distribution function (PDF) set~\cite{nnpdf} was used in the ME generation, in the parton shower, and in the simulation of multi-parton interactions (MPI). The ATLAS A14~\cite{ATL-PHYS-PUB-2014-021} set of tuned parameters (tune) was used for the parton shower and MPI, while hadronisation was modelled using the Lund string model~\cite{Andersson:1983ia,Sjostrand:1984ic}.
 
The \textsc{Sherpa} sample was generated using \textsc{Sherpa 2.1.1}. The matrix element calculation includes $2\to 2$ and $2\to 3$ processes at LO, merged using the CKKW method~\cite{Hoeche:2012yf}. The \textsc{Sherpa} parton shower~\cite{Catani:2001cc, Krauss:2002up} with $\pt$ ordering was used for the quark and gluon emissions. The matrix element renormalisation and factorisation scales for $2\to 2$ processes were set to the harmonic mean of the Mandelstam variables $s$, $t$ and $u$~\cite{mandelstam}, whereas the Catani--Marchesini--Webber (CMW)~\cite{CMWscale} scale was chosen for $2\to 3$ processes. The \textsc{CT14 NNLO}~\cite{ct14} PDF set was used for the matrix element calculation, while the parameters used for the modelling of the MPI and the parton shower were set according to the \textsc{CT10} tune~\cite{ct10}. The \textsc{Sherpa} sample made use of the dedicated Sherpa AHADIC model for hadronisation~\cite{Winter:2003tt}, which is based on the cluster fragmentation algorithm~\cite{Webber:1983if}.
 
Finally, two \textsc{Herwig 7} samples were generated using \textsc{Herwig 7.1.3} at next-to-leading order for the $2\to 2$ process. The matrix element was calculated using \textsc{Matchbox} \cite{Platzer:2011bc} with the \textsc{MMHT2014 NLO} PDF~\cite{mmht}. The renormalisation and factorisation scales were set to the $\pt$ of the leading jet. The first sample used an angle-ordered parton shower, while the second sample used a dipole-based parton shower. In both cases, the parton shower was interfaced to the matrix element calculation using the \textsc{MC@NLO} matching scheme~\cite{Frixione:2002ik}. The angle-ordered shower evolves on the basis of 1$\to $2  splittings using  a  generalised  angular  variable  and  employs  a  global recoil scheme once showering has terminated. The dipole-based shower uses 2$\to $3 splittings with Catani--Seymour kernels with an ordering in transverse momentum that enables it to perform recoils on an emission-by-emission basis. For both \textsc{Herwig 7} samples, the parameters that control the MPI and parton shower simulation were set according to the \textsc{MMHT2014} tune~\cite{mmht}, and the hadronisation was modelled by the cluster fragmentation algorithm.
 
The \textsc{Pythia 8}, \textsc{Sherpa} and \textsc{Herwig 7} samples were passed through the \textsc{Geant4}-based~\cite{Agostinelli:2002hh} ATLAS detector-simulation programs~\cite{SOFT-2010-01}, since they were used to unfold the measurements to the particle level as described in Section~\ref{sec:unfolding}. They were reconstructed and analysed with the same processing chain as the data. The generation of the simulated event samples includes the effect of multiple $pp$ interactions per bunch crossing, simulated using the \textsc{Pythia A3} tune~\cite{ATL-PHYS-PUB-2016-017}, and the effect of interactions from bunch crossings before or after the one containing the hard interaction.


\section{Jet reconstruction and event selection}
\label{sec:data_selection}
Jets are reconstructed using the anti-$k_t$ algorithm~\cite{antikt} with radius parameter $R=0.4$ using the \textsc{FastJet} program~\cite{Fastjet}. The inputs to the jet algorithm are particle-flow objects~\cite{pflow}, which make use of both the calorimeter and the inner-detector information to precisely determine the momenta of the input particles. The jet calibration procedure corrects for effects such as energy loss in inactive material, shower leakage, inefficiencies in energy clustering and jet reconstruction. This is performed using a simulation-based correction, in bins of $\eta$ and $\pt$, derived from the relation between the reconstructed jet energy and the energy of the corresponding particle-level jet, not including muons or non-interacting particles. The procedure also includes energy corrections for pile-up, and angular corrections. In a final step, an in situ calibration corrects for residual differences in the jet response and resolution between the MC simulation and the data using $\pt$-balance techniques for dijet, $\gamma+$jet, $\Zboson$+jet and multijet final states~\cite{JES1,JES2}.
 
Events were selected using a single-jet trigger~\cite{TRIG-2016-01} with a minimum $\pt$ threshold of 460~\GeV, to select high-$\pt$ jets while avoiding prescaled triggers. The selected jets are required to have $\pt > 60~\GeV$ and $|\eta| < 2.4$ to reject pile-up jets and reduce experimental uncertainties. In addition, jets are required to satisfy quality criteria that reject non-collision background and noise jets~\cite{ATLAS-CONF-2015-029}. More than 99\% of all signal jets pass these criteria. Events are required to have at least two selected jets. The two leading jets are further required to have the scalar sum of their transverse momenta, $H_{\mathrm{T}2} = p_{\mathrm{T}1} + p_{\mathrm{T}2} > 1~\TeV$. This ensures a trigger efficiency of $\approx$100\%. About 57.5 million events in data satisfy the selection criteria. The data are binned in ten intervals of $H_{\mathrm{T}2}$ in which the TEEC and ATEEC distributions are measured. The MC predictions are reweighted using the distribution of the average number of interactions per bunch crossing, so that the pile-up conditions in the simulation exactly match those in data.


\section{Unfolding to particle level}
\label{sec:unfolding}
To make detector-independent comparisons with particle-level parton shower MC predictions and fixed-order theoretical predictions, the measured distributions are corrected for distortions induced by the response of the ATLAS detector and associated reconstruction algorithms. The fiducial phase-space region is defined at particle level for all particles with a decay length $c\tau>$ 10 mm; these particles are referred to as `stable'. Particle-level jets are reconstructed using the anti-$k_t$ algorithm with $R=0.4$ using stable particles, excluding muons and neutrinos. The fiducial phase space closely follows the event selection criteria defined in Section~\ref{sec:data_selection}. Particle-level jets are required to have $\pt > 60~\GeV$ and $|\eta| < 2.4$. Events with at least two particle-level jets are considered. These events are also required to have $H_{\mathrm{T}2} > 1~\TeV$.
 
The unfolding is performed using an iterative algorithm based on Bayes theorem~\cite{bayes}, which takes into account inefficiencies, acceptance and resolution effects in the detector response that lead to bin migrations between the detector-level and particle-level phase spaces. For each observable, the method makes use of a transfer matrix, $M_{ij}$, obtained from MC simulation, that parameterises the probability of a jet pair generated in bin $i$ being reconstructed in bin $j$. The binning of the distributions is chosen as a compromise between maximising the number of bins while minimising migration between bins due to the resolution of the measured variables. This ensures that the purity from migrations is above 60\% in all regions of the phase space. The inverse problem can be written as a linear equation
 
\[
R_i = \sum_{j=1}^N \frac{\mathcal{E}_j}{\mathcal{P}_i} M_{ij} T_j.
\]
 
The quantities $R_i$ and $T_j$ are the contents of the detector-level distribution in bin $i$ and the contents of the particle-level distribution in bin $j$, respectively, while the factors $\mathcal{E}_j$ and $\mathcal{P}_i$ are efficiency and purity corrections, which are estimated by using the MC simulation.
 
The unfolding is performed separately in each bin of $H_{\mathrm{T}2}$ only considering the dependence on $\cos\phi$. The effect of migrations across different $H_{\mathrm{T}2}$ bins is small and is treated by the efficiency and purity corrections. It has been verified that using a two-dimensional transfer matrix, including $H_{\mathrm{T}2}$ and $\cos\phi$, gives results compatible within the statistical uncertainties.
 
Five iterations are necessary for the procedure to converge, i.e. for the relative difference between one iteration and the next, averaged over all bins of the distribution, to be below one per mille. The statistical uncertainties in both data and MC samples are propagated through the unfolding procedure by using pseudo-experiments. A set of 1000 replicas is considered for each measured distribution by applying a Poisson-distributed fluctuation around the nominal measured distribution. Each of these replicas is unfolded using a fluctuated version of the transfer matrix, which produces the corresponding set of 1000 replicas of the unfolded spectra. The statistical uncertainty is defined as the standard deviation of all replicas.
 
The results obtained by unfolding the data with \textsc{Pythia 8} are used to obtain the nominal differential cross-sections, whereas \textsc{Sherpa} and \textsc{Herwig 7} MC predictions are used to estimate the systematic uncertainty due to the model dependence, as discussed in Section~\ref{sec:systematics}.


\section{Experimental uncertainties}
\label{sec:systematics}
The dominant sources of systematic uncertainty in the measurements arise from knowledge of the jet energy scale and resolution and the modelling of the final state. The detector-level distributions are recomputed and unfolded for each systematic variation. The systematic uncertainties are then propagated through the unfolding via their impact on the transfer matrices, the purity and the efficiency. A detailed description of the systematic uncertainties and their typical values for the TEEC and ATEEC functions is given below:
\begin{itemize}
\item{} Jet Energy Scale and Resolution: The jet energy scale (JES) and jet energy resolution (JER) uncertainties are estimated as described in Ref.~\cite{JES1,JES2}. The JES uncertainties consist of a set of 113 independent components, which depend on the jet $\pt$ and $\eta$. These components account for different effects, including the different response to quarks and gluons and the flavour composition of the sample. Sources related to the in situ methods for the determination of JES~\cite{JES1,JES2} are also included. The total JES uncertainty in the $\pt$ value of individual jets varies from approximately 2\% at $\pt$~=~60~\GeV\ to below 1\% for $\pt > 500~\GeV$, with a mild dependence on $\eta$. The JER uncertainty is estimated by using 34 independent components, arising from the determination of the JER noise term and the different in situ methods~\cite{JES2}. The effect of the total JER uncertainty is evaluated by smearing the energy of the jets in the MC simulation by about 2\% at $\pt$ = 60~\GeV\ to about 0.5\% for $\pt$ of several hundred \GeV. Each JES and JER uncertainty component is independently propagated by varying the energy and $\pt$ of each jet by one standard deviation. The total uncertainty in the TEEC distribution ranges from a maximum of 2\% for $1~\TeV < H_{\mathrm{T2}} < 1.2~\TeV$ to 1.5\% for $H_{\mathrm{T}2} > 3.5~\TeV$ while, for the ATEEC, it is below 1\% in the whole phase space of the analysis.
 
\item{} Jet Angular Resolution: The jet angular resolution (JAR) uncertainty is conservatively estimated by smearing the azimuthal coordinate $\phi$ of the jets by their resolution in MC simulation. The $\phi$ variations are performed with the $p_{\textrm{T}}$ component of the jets held constant. The value of the JAR uncertainty is below 0.5\% for both the TEEC and ATEEC distributions in all regions of phase space.
 
\item{} Unfolding: The mismodelling of the data by the MC simulation is accounted for as an additional source of uncertainty. This is assessed by reweighting the particle-level distributions so that the detector-level distributions predicted by the MC samples match those in the data. The modified detector-level MC distributions are then unfolded using the method described in Section~\ref{sec:unfolding}. The difference between the modified particle-level MC distribution and the original particle-level MC distribution is taken as the uncertainty. This uncertainty is negligible for both the TEEC and ATEEC in each of the phase space regions of the analysis.
 
\item{} Monte Carlo Modelling: The modelling uncertainty is estimated by comparing the distributions obtained by using \textsc{Pythia 8}, \textsc{Sherpa} and \textsc{Herwig 7} in the unfolding procedure. The envelope of the differences between the estimated cross-sections defines the systematic uncertainty. The value of this uncertainty for the TEEC varies from 1.5\% to 2\%, depending on the phase space region while, for the ATEEC, it is always below 0.5\% and has an impact only for $\cos\phi < -0.95$.
\end{itemize}
 
A breakdown of the relative systematic uncertainties in the measurement of the TEEC and ATEEC distributions is shown in Figure~\ref{fig:syst}.
 
\begin{figure}[H]
\centering
\subfloat[]{\includegraphics[width=0.49\textwidth]{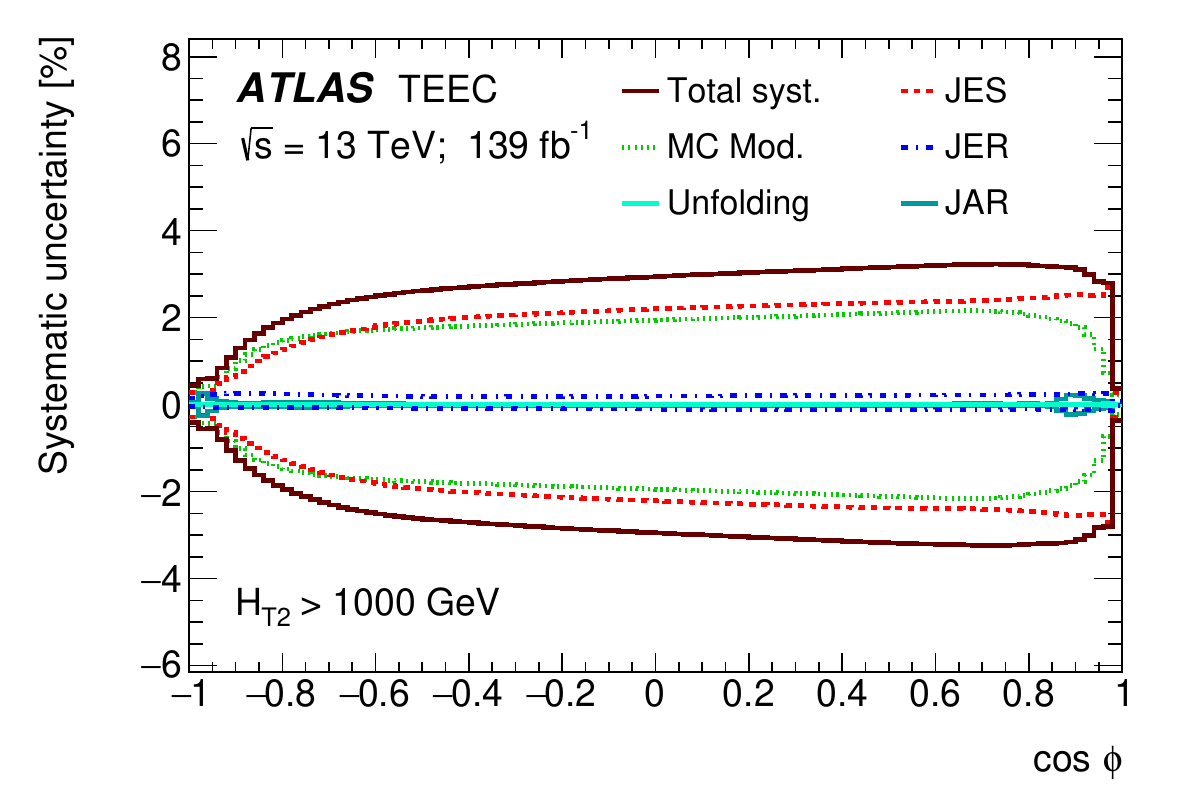}}
\subfloat[]{\includegraphics[width=0.49\textwidth]{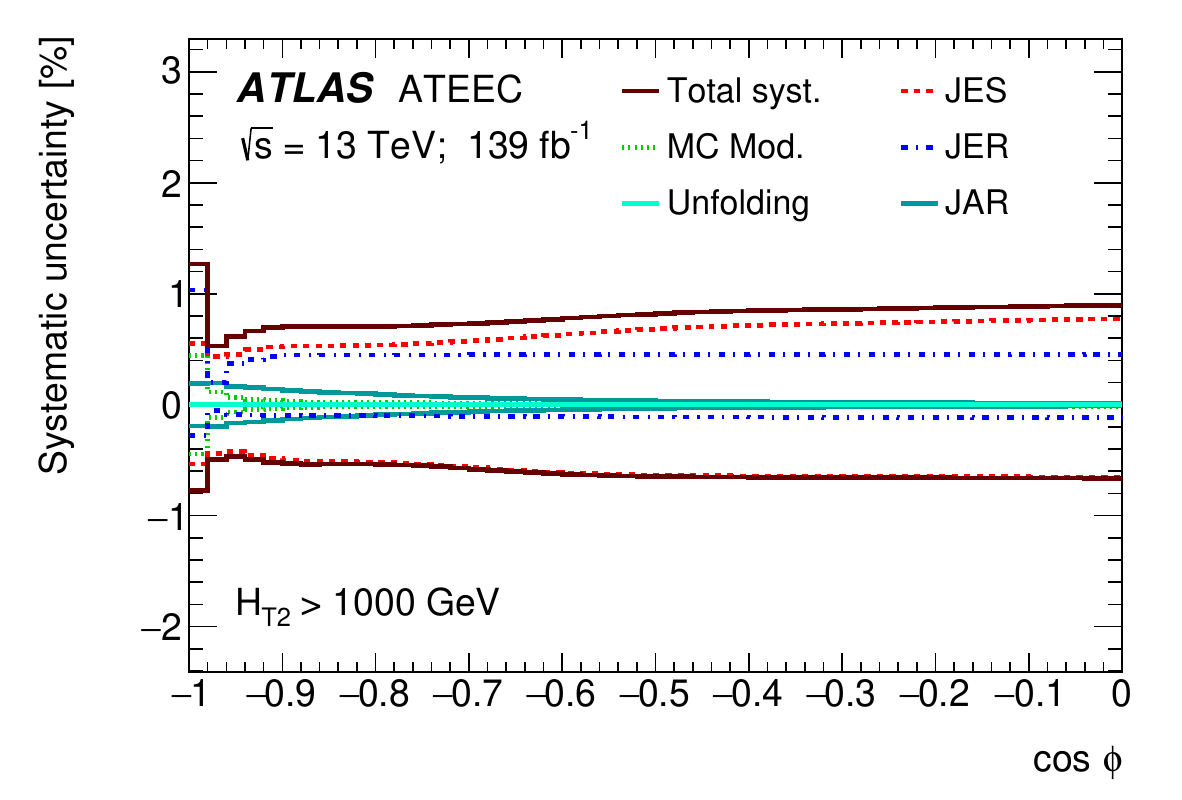}}
\caption{Breakdown of the systematic uncertainties as a function of (a) the TEEC and (b) the ATEEC for the inclusive $H_{\mathrm{T}2}$ sample, including jet energy scale and resolution, jet angular resolution and uncertainties in the unfolding procedure. Due to the normalisation of the TEEC distribution, the systematic uncertainties show a sharp change in magnitude close to $\cos\phi \simeq 1.0$ for the TEEC and to $\cos\phi\simeq-1.0$ for the ATEEC.}
\label{fig:syst}
\end{figure}

The total systematic uncertainty is estimated by adding the effects previously listed in quadrature. In addition, the statistical uncertainty in the data and MC simulation is propagated to the differential cross-sections through the unfolding procedure using pseudo-experiments to properly take into account the statistical correlations between bins. The pseudo-experiments are also used to estimate the statistical component of each systematic uncertainty. This statistical component is mitigated using the Gaussian Kernel smoothing technique~\cite{PERF-2012-01}. The values provided above are quoted after application of this procedure. The impact of these sources of uncertainty depends on the region of phase space. Overall, the JES uncertainty dominates both the TEEC and ATEEC measurements, while the MC modelling and JER uncertainties are also important for the TEEC and ATEEC measurements, respectively.


\section{Experimental results}
The unfolded TEEC and ATEEC distributions are presented and compared with the MC predictions described in Section~\ref{sec:MC_simulation}. Figure~\ref{fig:resultsMC1} shows this comparison for the TEEC and ATEEC distributions obtained in the inclusive region $H_{\mathrm{T}2} > 1~\TeV$. Additionally, Figure~\ref{fig:resultsMC2} shows this comparison for both the TEEC and ATEEC for two different bins of $H_{\mathrm{T}2}$.
 
\begin{figure}[H]
\centering
\subfloat[]{\includegraphics[width=0.78\textwidth]{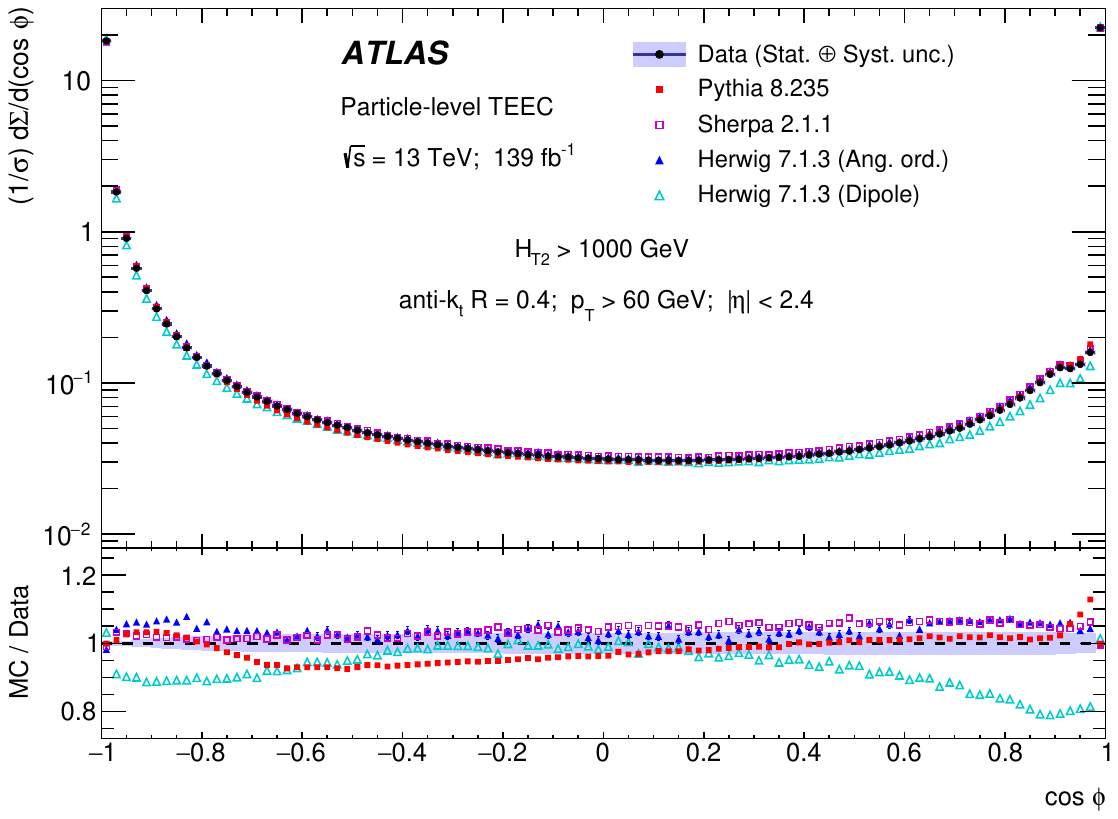}}\\
 
\subfloat[]{\includegraphics[width=0.78\textwidth]{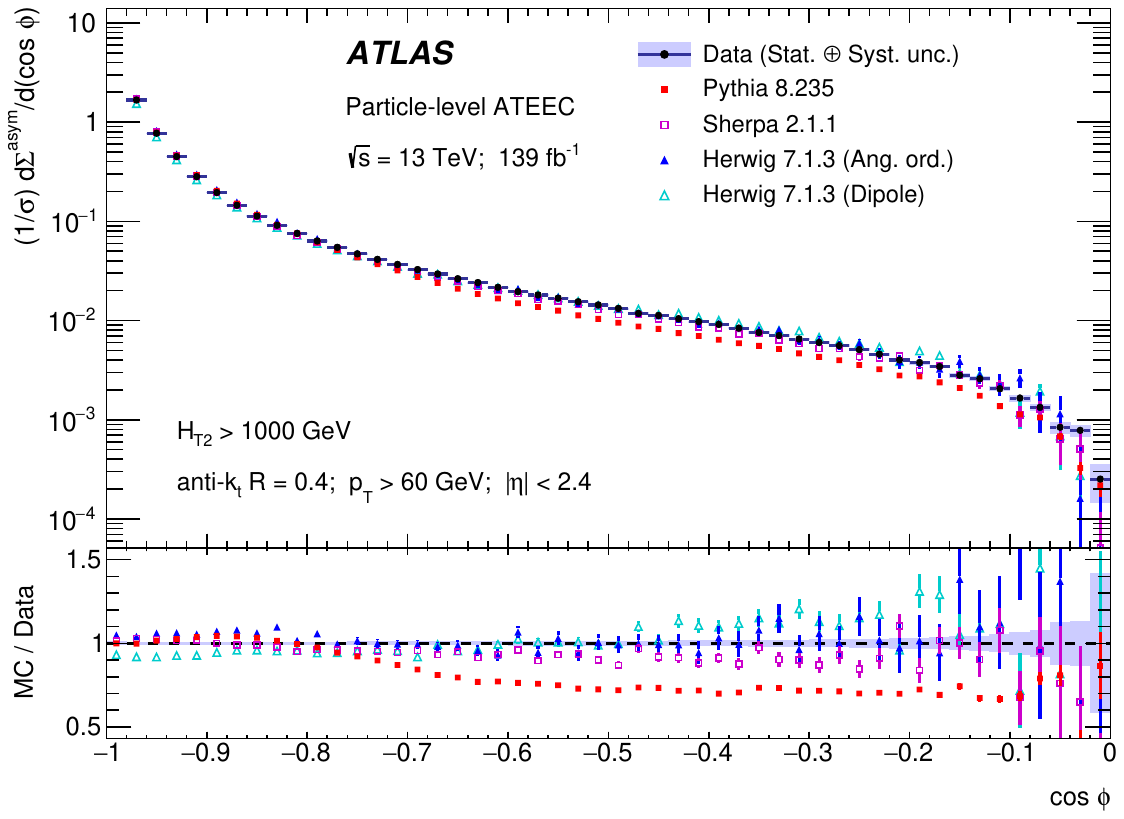}}
\caption{Comparison between the data and the MC predictions for (a) the TEEC and (b) the ATEEC for the inclusive $H_{\mathrm{T}2}$ bin. The data error bands contain both statistical and systematic components, summed in quadrature. The lower panels show the ratios of the MC predictions to the data, including the statistical uncertainty on the MC predictions. The value of the first bin of the ATEEC distribution is negative and, therefore, not represented in logarithmic scale.}
\label{fig:resultsMC1}
\end{figure}
 
\begin{figure}[H]
\centering
\subfloat[]{\includegraphics[width=0.44\textwidth]{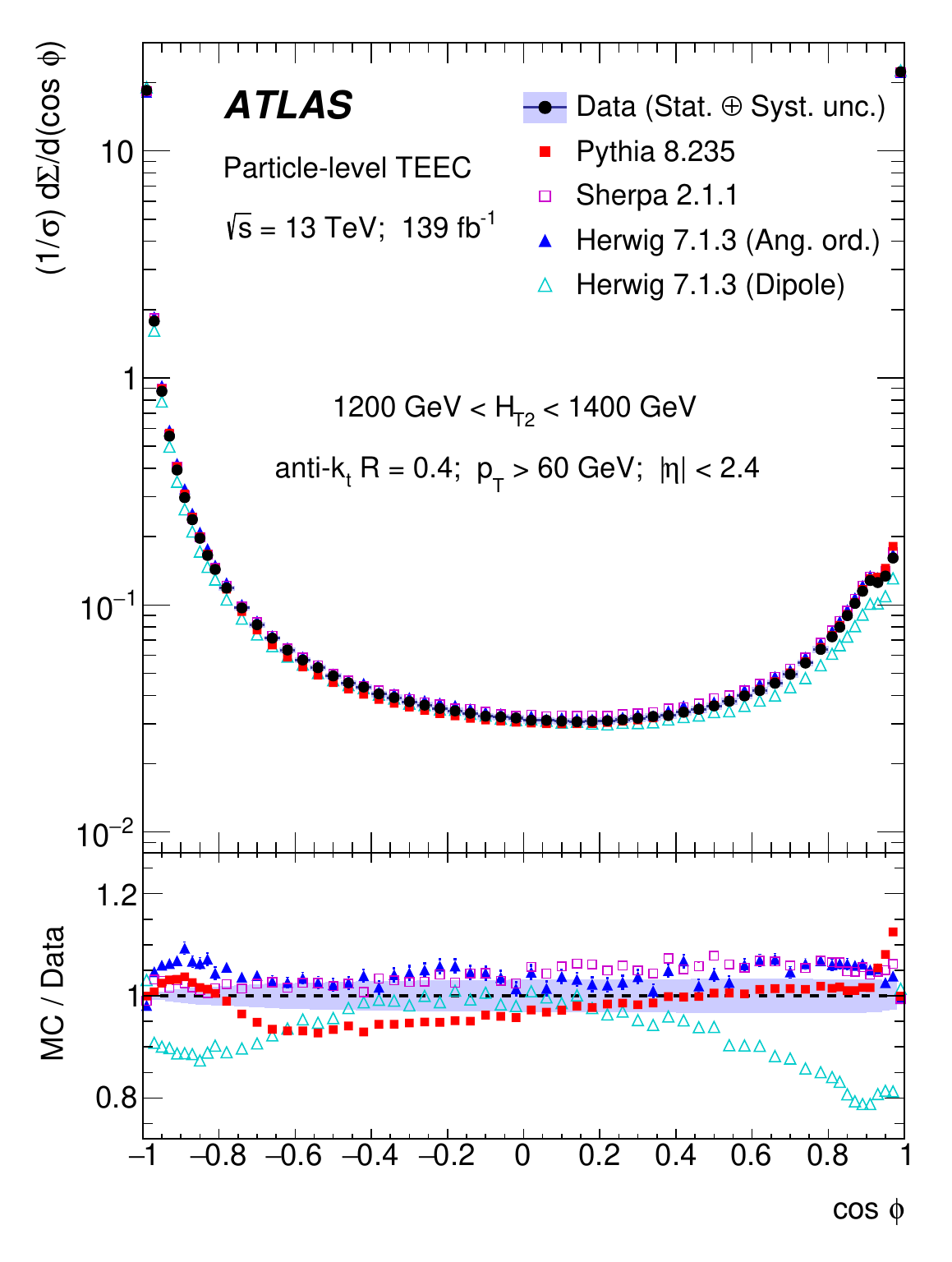}}
\subfloat[]{\includegraphics[width=0.44\textwidth]{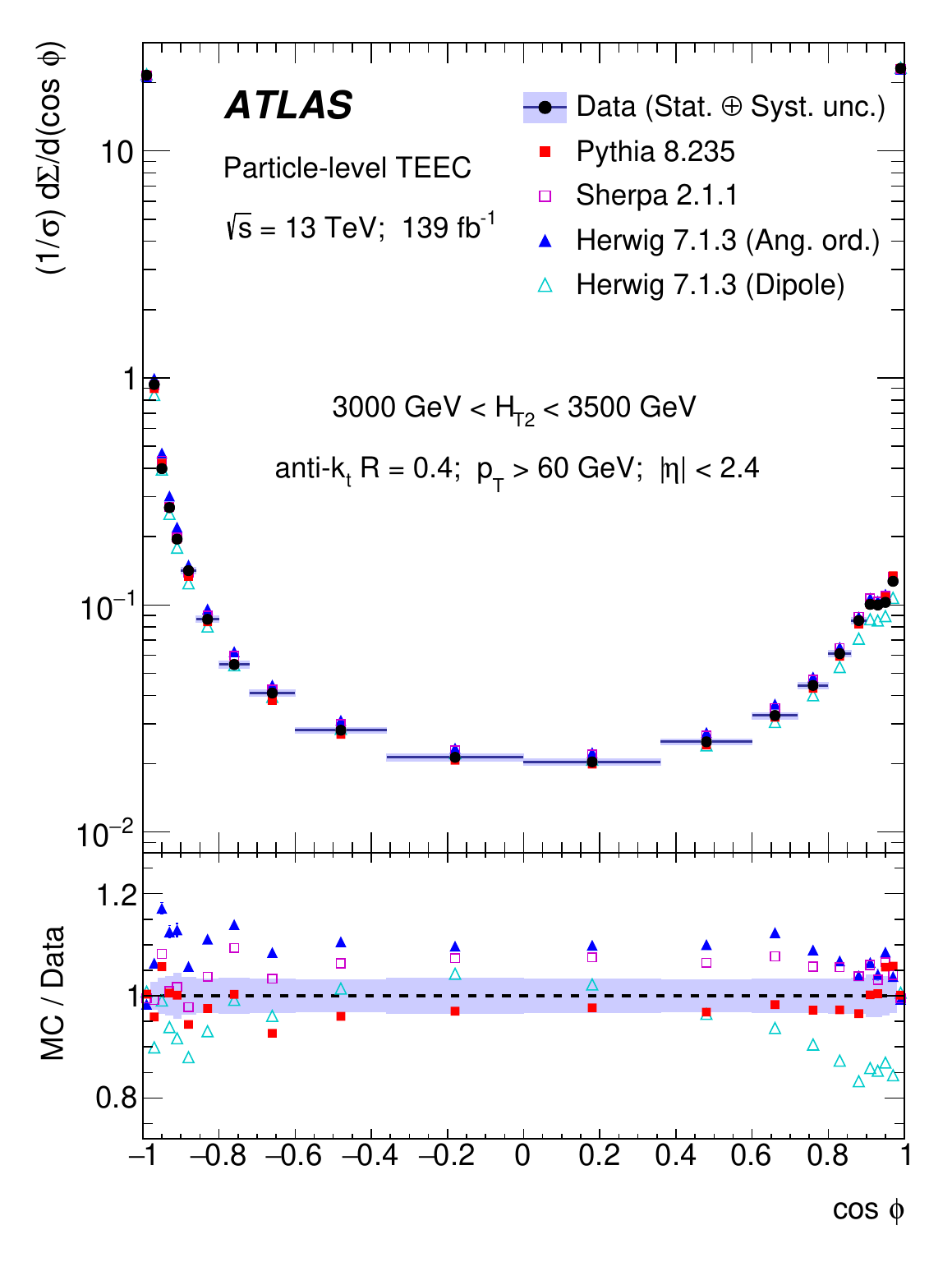}}
 
\subfloat[]{\includegraphics[width=0.44\textwidth]{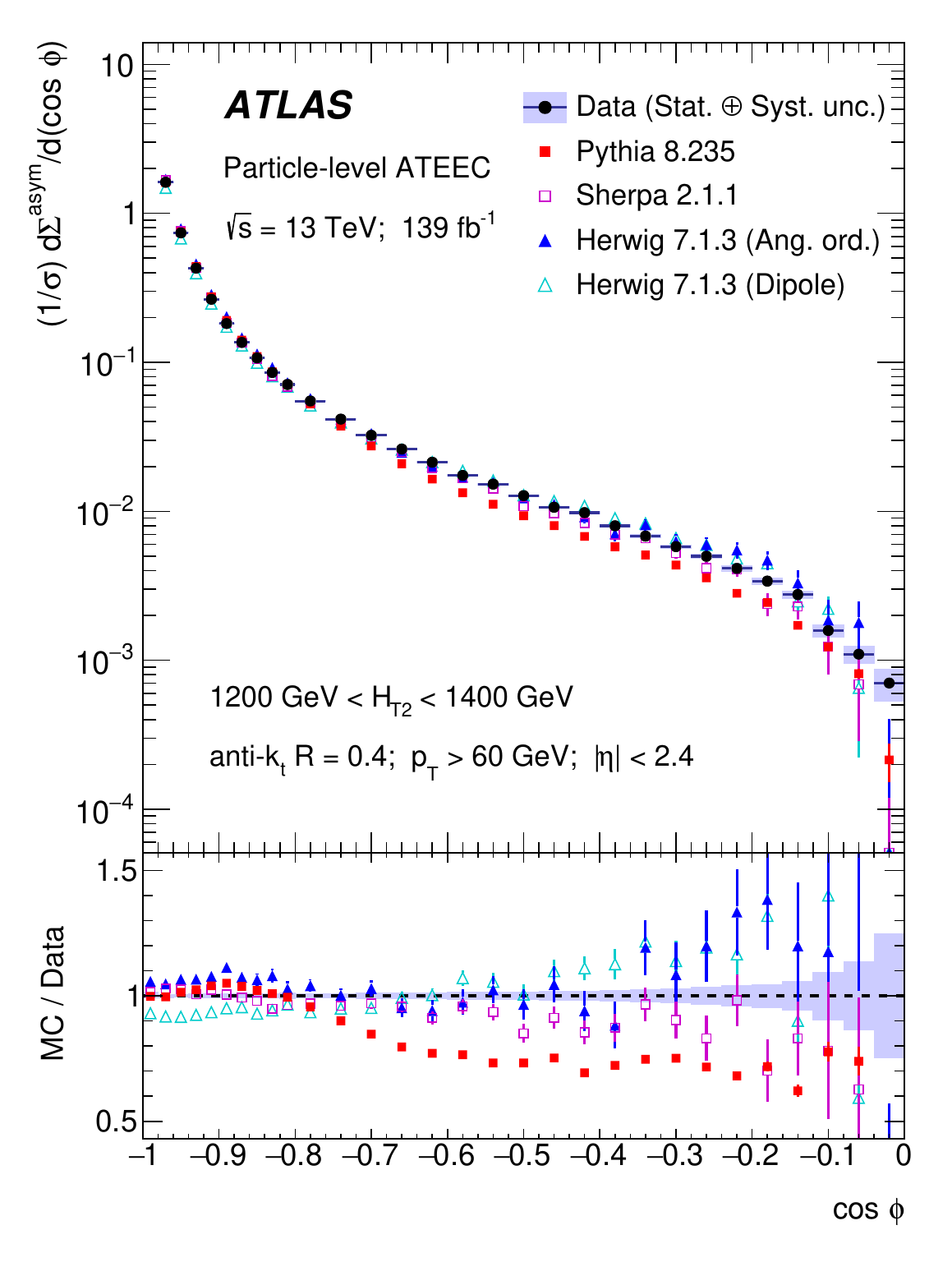}}
\subfloat[]{\includegraphics[width=0.44\textwidth]{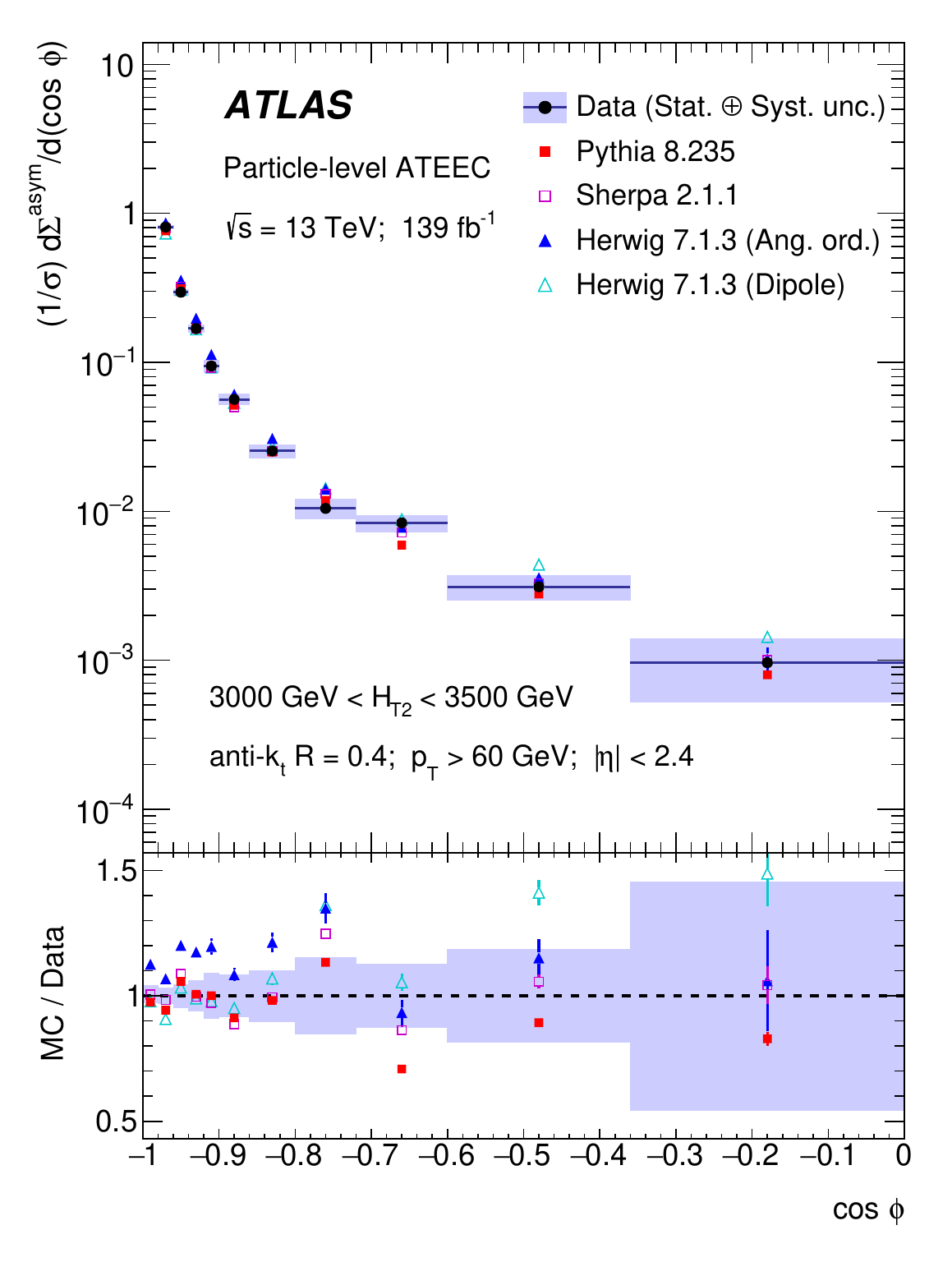}}
\caption{Comparison between the data and the MC expectations for (a, b) the TEEC and (c, d) the ATEEC for two selected regions of $H_{\mathrm{T}2}$. The data error bands contain both statistical and systematic components, summed in quadrature. The lower panels show the ratios of the MC predictions to the data, including the statistical uncertainty on the MC predictions. The values of the first bins of the ATEEC distributions are negative and, therefore, not represented in logarithmic scale.}
\label{fig:resultsMC2}
\end{figure}
The TEEC and ATEEC distributions presented in Figures~\ref{fig:resultsMC1} and~\ref{fig:resultsMC2} show the features observed in previous results~\cite{STDM-2014-03, STDM-2016-10}. The TEEC shows two peaks arising from back-to-back ($\cos\phi = -1$) and collinear ($\cos\phi = 1$) configurations, together with a central plateau arising from the wide-angle radiation. The effect of the jet radius $R$ is seen as a kink in the TEEC distributions at $\cos\phi \simeq 0.92$. The ATEEC exhibits a steep fall-off of several orders of magnitude between $\cos\phi = -1$ and $\cos\phi = 0$.
 
At low values of $H_{\mathrm{T}2}$, the best description of the TEEC distributions is achieved by both \textsc{Sherpa} and \textsc{Herwig 7} with the angle-ordered parton shower. The dipole parton shower in \textsc{Herwig 7} does not describe the data well, in particular in the region $|\cos\phi| > 0.4$, where discrepancies of up to 20\% are observed close to the edges of the TEEC distribution. \textsc{Pythia 8} shows discrepancies compared with data for $-0.7 < \cos\phi < -0.2$, where differences of $\mathcal{O}(5\%-10\%)$ are observed. A good description of the ATEEC distribution is achieved by both \textsc{Sherpa} and \textsc{Herwig 7}, while \textsc{Pythia 8} underestimates the measured distribution for $\cos\phi > -0.7$ and the \textsc{Herwig 7} sample matched to the dipole parton shower underestimates the values of the ATEEC for $\cos \phi < -0.7$.
 
For large values of $H_{\mathrm{T}2}$, \textsc{Pythia 8} gives the best description of the TEEC data, while both \textsc{Sherpa} and \textsc{Herwig 7} tend to overestimate the height of the central plateau. The dipole parton shower in \textsc{Herwig 7} shows a similar behaviour to the low $H_{\mathrm{T}2}$ region, with the larger discrepancies observed near the edges of the TEEC distribution, although limited to the region $|\cos\phi| > 0.7$. For the ATEEC, \textsc{Pythia 8} and \textsc{Sherpa} give a good description of the data, while the \textsc{Herwig 7} sample matched to the angle-ordered parton shower overestimates the values of the ATEEC for $\cos\phi < -0.7$. The cancellation of discrepancies in $\phi$-symmetric regions between the data and the dipole shower in \textsc{Herwig 7} brings a reasonable agreement for the ATEEC in the high energy regime.


\section{Theoretical predictions and uncertainties}
The theoretical predictions for the TEEC and ATEEC functions are calculated using pQCD at LO, NLO and NNLO in powers of the strong coupling constant, $\alpha_{\mathrm{s}}(\mu_\mathrm{R})$, as discussed in Ref.~\cite{mitov2022, mitov2023}. The calculations make use of \textsc{OpenLoops2}~\cite{Buccioni:2019sur}, \textsc{FivePointAmplitudes}~\cite{fivepoint} and \textsc{PentagonFunctions++}~\cite{pentagonfunctions} for the evaluation of the scattering amplitudes. A mixed flavour scheme is assumed. On the one hand, the effect of the top quark is taken into account when evaluating the radiative corrections to the gluon propagator. Thus, the solution to the RGE includes six active flavours by considering different values of $\Lambda_{\mathrm{QCD}}$ when crossing the corresponding mass thresholds. On the other hand, the perturbative corrections to the three-jet cross-section assume $n_{\mathrm{f}} = 5$ massless quark flavours, thus neglecting the effect of the top quark in both the virtual and real contributions. The contribution from real top-quark production is estimated to be below 0.3\% for the phase space relevant to this analysis.
 
At $\mathcal{O}(\alpha_s^5)$ the numerator in Eq.~\ref{eq:teecDef} entails the calculation of $2\to3$ subprocesses at NNLO, $2\to 4$ at NLO and $2\to 5$ at LO. In order to avoid the triple-collinear singularities appearing in the latter, the angular range is restricted to $|\cos \phi| < 0.92$. This avoids calculating the three-loop virtual corrections to the $2\to 2$ subprocesses. Thus, with this azimuthal cut, the denominator in Eq.~\ref{eq:teecDef} includes the $2\to 2$, $2\to 3$ and $2\to 4$ subprocesses up to, and including, $\mathcal{O}(\alpha_s^4)$ corrections.
 
The value of $\alpha_{\mathrm{s}}(\mu_\mathrm{R})$ that enters in the partonic matrix-element calculation is obtained  by evolving the same $\alpha_{\mathrm{s}}(m_\Zboson)$ value as used to determine the PDF set under consideration\@. The NNLO PDF sets \textsc{MMHT2014}~\cite{mmht}, \textsc{NNPDF 3.0}~\cite{nnpdf3} and CT14~\cite{ct14}, as provided in the LHAPDF package~\cite{LHAPDF}, are convolved with the partonic cross-sections.
 
The NNLO corrections are calculated using $\mathcal{O}(10^{13})$ events. They include real--real, real--virtual and virtual--virtual finite terms, as well as single- and double-unresolved terms and the finite remainder~\cite{NNLOscheme1,NNLOscheme2}. For illustrative purposes, Figure~\ref{fig:nnloDiagrams} shows sample diagrams corresponding to real--real, real--virtual and virtual--virtual contributions to the $gg\to \mathrm{jjj}$ process at NNLO.
 
\begin{figure}[H]
\centering
\subfloat[]{\includegraphics[width=0.23\textwidth]{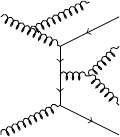}}
\hspace{1.2cm}
\subfloat[]{\includegraphics[width=0.23\textwidth]{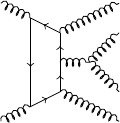}}
\hspace{1.2cm}
\subfloat[]{\includegraphics[width=0.23\textwidth]{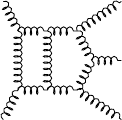}}
\caption{Feynman diagrams contributing to the $gg\to \mathrm{jjj}$ process at NNLO accuracy in pQCD\@. These particular configurations contribute to the (a) real--real, (b) real--virtual, and (c) virtual--virtual finite terms.}
\label{fig:nnloDiagrams}
\end{figure}
For each event, the renormalisation and factorisation scales are set to the scalar sum of the transverse momenta of all final-state partons, $\mu_R = \mu_F = \hat{H}_{\mathrm{T}}$~\cite{CERN-TH-2018-159, mitov2022, mitov2023}.
 
The pQCD predictions for the TEEC and ATEEC functions are obtained using jets clustered from final-state partons as inputs to the anti-$k_t$ algorithm with $R=0.4$~\cite{antikt}. Therefore, non-perturbative correction factors are derived to account for hadronisation and underlying event effects on the observables. These correction factors are calculated as the ratio of the MC predictions including both hadronisation and underlying event effects to the MC predictions in which these effects are not included. The central values for these correction factors are calculated using the A14 tune in \textsc{Pythia~8}~\cite{ATL-PHYS-PUB-2014-021}, and are close to unity to within 0.5\%, except for $\cos\phi > 0.86$ where they go down to as far as 0.95.
 
The uncertainties in the theoretical predictions arise from three sources: contributions from higher-order corrections estimated using scale variations, the uncertainties in the PDF and the uncertainties in the non-perturbative corrections to the pQCD predictions.
\begin{itemize}
\item{} The scale uncertainties are evaluated by considering independent variations of the renormalisation and factorisation scales by a factor of two, excluding those in which $\mu_\mathrm{R}$ and $\mu_\mathrm{F}$ are varied in opposite directions. As in previous analyses~\cite{STDM-2014-03, STDM-2016-10}, the scale variations are treated as correlated between the three-jet and two-jet cross sections, respectively in the numerator and denominator of Eq.~\ref{eq:teecDef}. The uncertainty in the distributions is defined as the envelope of the six resulting variations. Overall, the scale uncertainty bands are asymmetric and larger for the lower part of the TEEC and ATEEC distributions. The value of the scale uncertainties are below 2\% in all regions of phase space, while they account for up to 6\% at NLO\@. Thus, this source of uncertainty is reduced by a factor of three by the NNLO corrections. This is the dominant theoretical uncertainty for both TEEC and ATEEC.
\item{} The PDF uncertainties are obtained by considering the set of eigenvectors / replicas provided by each PDF group~\cite{mmht,nnpdf3,ct14}, and are propagated to the predicted distributions using the corresponding recommendations from each particular PDF set. Due to the large computing time required for the NNLO calculations, the PDF variations are calculated at NLO and extrapolated to the NNLO predictions using their relative values. The resulting uncertainties are below 1\% for the low $H_{\mathrm{T2}}$ bins, and can reach up to 2\% for the higher $H_{\mathrm{T2}}$ bins.
\item{} The uncertainties in the non-perturbative corrections are evaluated by considering the envelope of the differences between five different predictions and the \textsc{Pythia 8} A14 tune~\cite{ATL-PHYS-PUB-2014-021}. These predictions were generated using \textsc{Pythia 8.235} and \textsc{Herwig 7.2.1} with the angle-ordered parton shower. They include the \textsc{4C}~\cite{Tune4C} and \textsc{Monash}~\cite{Skands:2014pea} \textsc{Pythia 8} tunes, and the \textsc{Herwig 7} default, soft and baryonic reconnection tunes~\cite{herwigTunes}. The value of this uncertainty is below 1\% in the phase-space region $|\cos\phi| < 0.92$, where $\alpha_{\mathrm{s}}$ is determined.
\end{itemize}
The relative values of the theoretical uncertainties for the TEEC and ATEEC distributions are shown in Figure~\ref{fig:theoSys} for the inclusive $H_{\mathrm{T2}}$ bin. The effect of varying $\alpha_s(m_\Zboson)$ within the range 0.117 -- 0.119 is also shown for illustrative purposes.
\begin{figure}[H]
\subfloat[]{\includegraphics[width=0.49\textwidth]{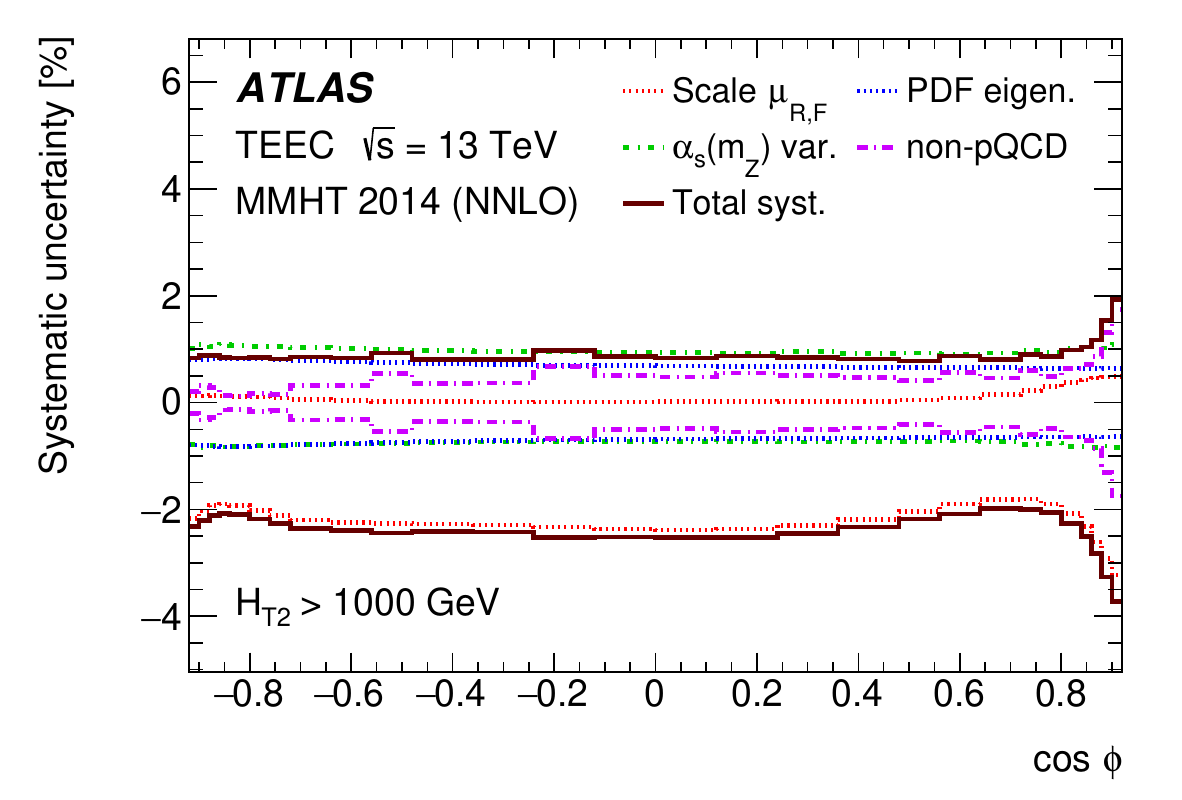}}
\subfloat[]{\includegraphics[width=0.49\textwidth]{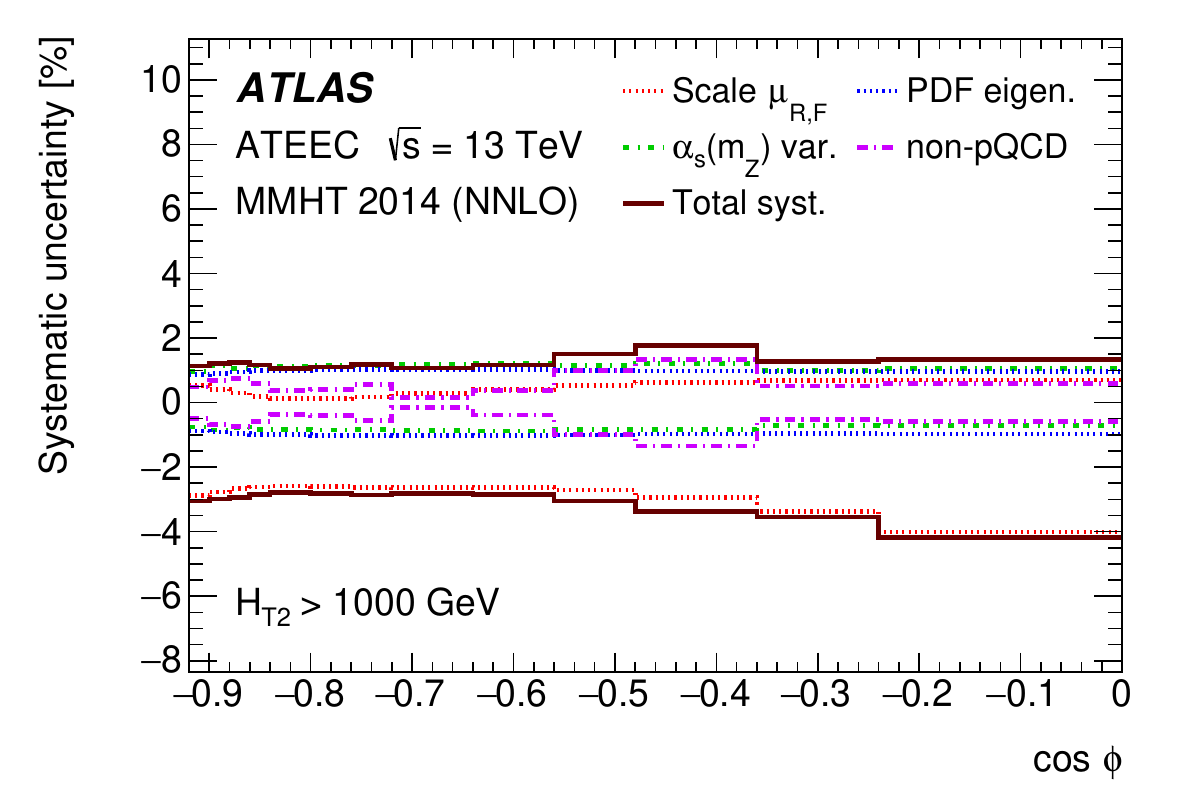}}
\caption{Theoretical uncertainties in the NNLO pQCD prediction for (a) the TEEC and (b) the ATEEC distributions for $H_{\mathrm{T}2} > 1$~\TeV. The predictions use the \textsc{MMHT2014} PDF\@. The effect of a variation of $\alpha_s(m_\Zboson)$ from 0.118 by $\pm 0.001$ is also shown for illustration.}
\label{fig:theoSys}
\end{figure}
Figures~\ref{fig:ratioTEEC} and~\ref{fig:ratioATEEC} show the ratios of the data, as well as of the LO and NNLO theoretical predictions to the NLO predictions for the TEEC and ATEEC, respectively. The description of the data provided by the NNLO pQCD predictions is very good, improving with respect to the NLO prediction. In these figures, the distributions are rebinned to reduce statistical fluctuations in the theoretical predictions that, despite the high statistics, can be present due to insufficient precision on the cancellation of the different terms. The reduction of the scale uncertainties is made evident from these figures, as well as the improvement in the description. In particular, the region with $\cos\phi > 0.80$, shows a large improvement for all $H_{\mathrm{T2}}$ bins. For the higher $H_{\mathrm{T2}}$ bins, and for the \textsc{MMHT2014} PDF, the NNLO theoretical predictions are slightly above the data. Since the \textsc{CT14} PDF shows a better behaviour in this region of the phase space, this effect may be due to the limited accuracy of the PDF determinations at high values of the Bjorken variable $x$. The results obtained using the \textsc{CT18} PDF set agree well with those for \textsc{MMHT2014}.
 
\begin{figure}[H]
\vspace{1.8cm}
\centering
\includegraphics[width=0.955\textwidth]{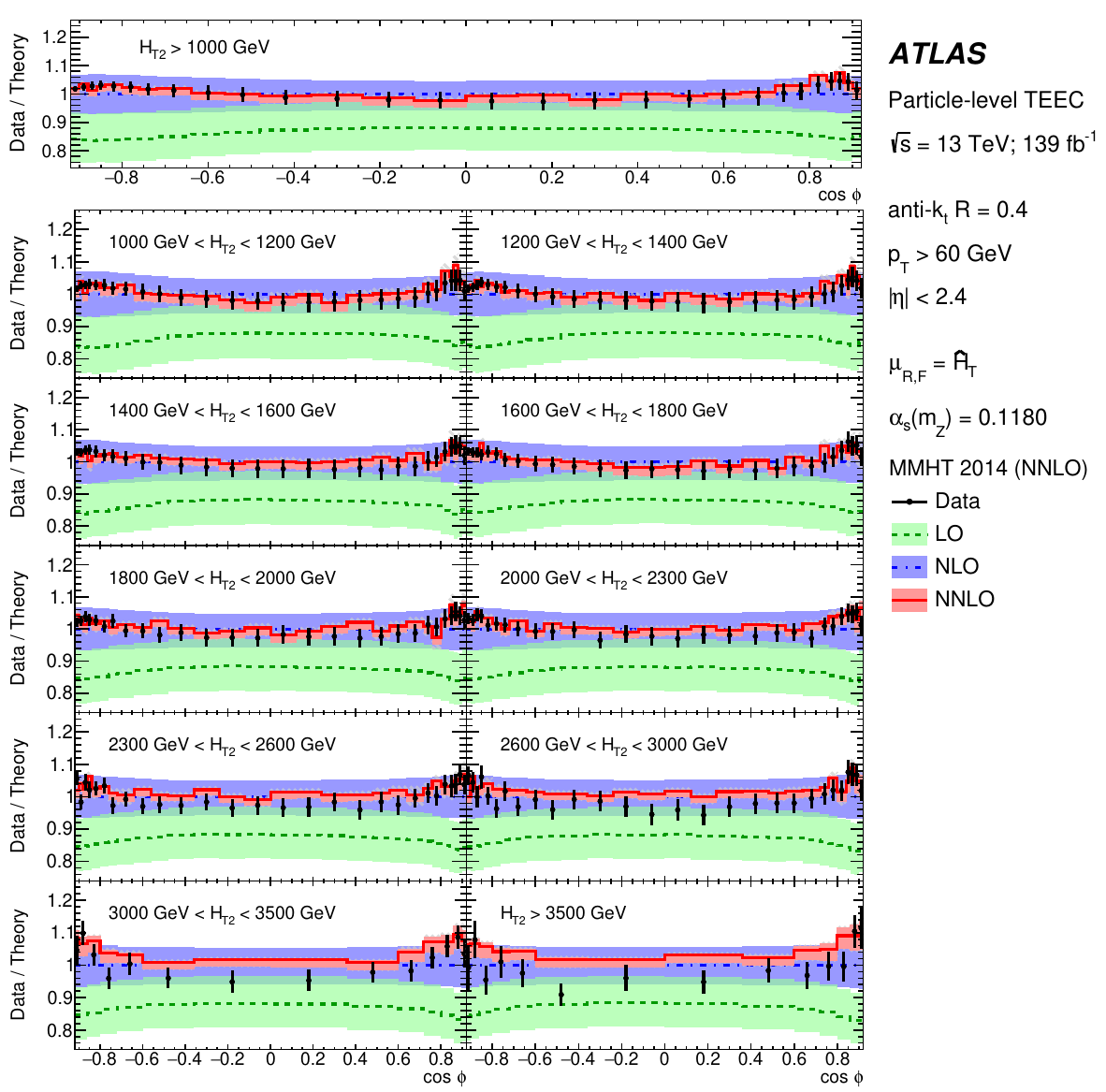}
\caption{Ratios of the theoretical predictions for the TEEC functions at LO and NNLO to the NLO calculations, together with the ratios of the data to NLO predictions. The hatched band, where visible, shows the statistical uncertainty in the NNLO prediction. The predictions use the \textsc{MMHT2014} PDF, where the value of the strong coupling constant is set to $\alpha_{\mathrm{s}}(m_\Zboson) = 0.1180$. The uncertainty bands correspond to the scale uncertainties for each perturbative order.}
\label{fig:ratioTEEC}
\end{figure}
\begin{figure}[H]
\vspace{1.8cm}
\centering
\includegraphics[width=0.955\textwidth]{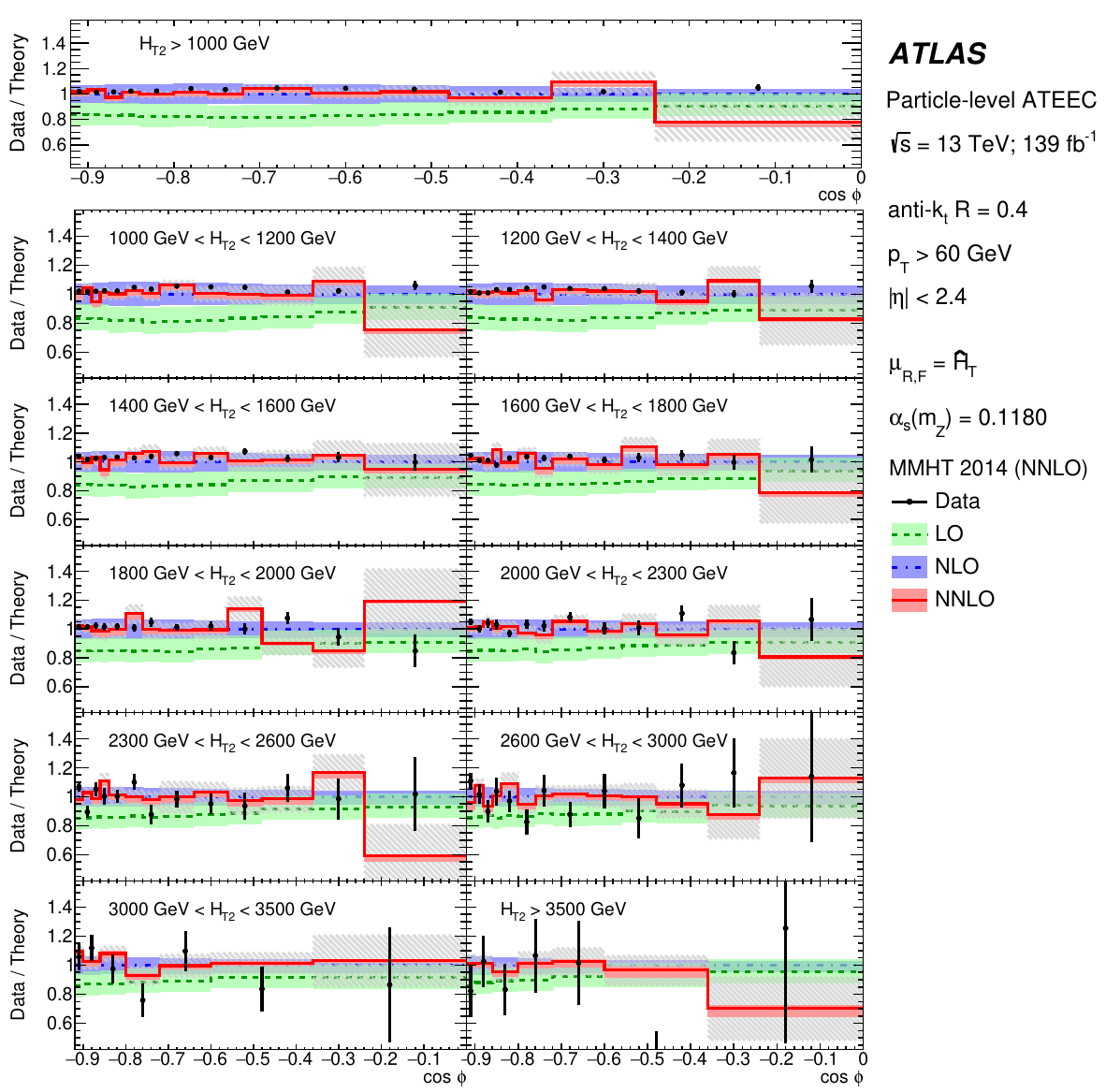}
\caption{Ratios of the theoretical predictions for the ATEEC functions at LO and NNLO to the NLO calculations, together with the ratios of the data to NLO predictions. The hatched band, where visible, shows the statistical uncertainty in the NNLO prediction. The predictions use the \textsc{MMHT2014} PDF, where the value of the strong coupling constant is set to $\alpha_{\mathrm{s}}(m_\Zboson) = 0.1180$. The uncertainty bands correspond to the scale uncertainties for each perturbative order.}
\label{fig:ratioATEEC}
\end{figure}
\clearpage


\section{Determination of the strong coupling constant}
The strong coupling constant at the scale of the pole mass of the $\Zboson$ boson, $\alpha_{\mathrm{s}}(m_\Zboson)$ can be determined from the comparison of the data with the theoretical predictions by considering the following $\chi^2$ function
\begin{equation}
\chi^2(\alpha_{\mathrm{s}},\vec{\lambda}) = \sum_{\mathrm{bins}}\frac{(x_i-F_i(\alpha_{\mathrm{s}},\vec{\lambda}))^2}{\Delta x_i^2 + \Delta \xi_i^2} + \sum_{k}\lambda_k^2,
\label{eq:chi2a}
\end{equation}
where the theoretical predictions are varied according to
\begin{equation}
F_{i}(\alpha_{\mathrm{s}},\vec{\lambda}) = \psi_{i}(\alpha_{\mathrm{s}})\left(1+\sum_k\lambda_k\sigma_k^{(i)}\right).
\label{eq:chi2b}
\end{equation}
In Eq.~(\ref{eq:chi2a}) and Eq.~(\ref{eq:chi2b}), $\alpha_{\mathrm{s}}$ stands for $\alpha_{\mathrm{s}}(m_\Zboson)$; $x_i$ is the value of the $i$-th point of the distribution as measured in data, while $\Delta x_i$ is its statistical uncertainty. The statistical uncertainty in the theoretical predictions dominates over the statistical uncertainty in data, and is also included as $\Delta \xi_i$, while $\sigma_k^{(i)}$ is the relative value of the $k$-th correlated source of systematic uncertainty in bin $i$.
 
This technique takes into account the correlations between the different sources of systematic uncertainty discussed in Section~\ref{sec:systematics} by introducing the nuisance parameters $\{\lambda_k\}$, one for each source of experimental uncertainty, each of which follow a standard normal distribution. Thus, the minimum of the $\chi^2$ function defined in Eq.~(\ref{eq:chi2a}) is found in a 150-dimensional space, in which 149 correspond to nuisance parameters $\left\{\lambda_i\right\}$ and one to $\alpha_{\mathrm{s}}(m_\Zboson)$. The uncertainty due to the choice of the MC model in the unfolding is treated by performing alternative fits in which the data distribution is unfolded using different MC models, including \textsc{Herwig 7} and \textsc{Sherpa}. The envelope of the $\alpha_s$ values obtained for each model is used as the systematic uncertainty.
 
The method also requires an analytical expression for the dependence of each observable on the strong coupling constant, which is given by $\psi_{i}(\alpha_{\mathrm{s}})$ for bin $i$. For each PDF set, the corresponding $\alpha_{\mathrm{s}}(m_\Zboson)$ variation range is considered and the theoretical prediction is obtained for each value of $\alpha_{\mathrm{s}}(m_\Zboson)$. The functions $\psi_{i}(\alpha_{\mathrm{s}})$ are then obtained by fitting the predicted values of the TEEC (ATEEC) in each $(H_{\mathrm{T}2}, \cos\phi)$ bin to a third-order polynomial in $\alpha_s$.
 
For both the TEEC and ATEEC functions, the fits to extract $\alpha_{\mathrm{s}}(m_\Zboson)$ are repeated separately for each $H_{\mathrm{T}2}$ interval, thus determining a value of $\alpha_{\mathrm{s}}(m_\Zboson)$ for each energy bin. The theoretical uncertainties are determined by shifting the theory distributions by each of the uncertainties separately, recalculating the functions $\psi_i(\alpha_{\mathrm{s}})$ and determining a new value of $\alpha_{\mathrm{s}}(m_\Zboson)$. The uncertainty is determined by taking the difference between this value and the nominal one. The uncertainty due to the choice of the MC tune in the non-perturbative correction is estimated by repeating the fit with the theory distributions corrected with the other tunes, and taking the envelope relative to the nominal fit.
 
Each of the fitted values of $\alpha_{\mathrm{s}}(m_\Zboson)$, which are obtained in the $\overline{\mbox{MS}}$ subtraction scheme, is then evolved to the corresponding measured scale using the NNLO solution to the RGE\@. When evolving $\alpha_{\mathrm{s}}(m_\Zboson)$ to $\alpha_{\mathrm{s}}(Q)$, the appropriate matching conditions for the strong coupling constant at the $n_{\mathrm{f}}=5$ and $n_{\mathrm{f}}=6$ flavour thresholds are applied so that $\alpha_{\mathrm{s}}(Q)$ is a continuous function across quark thresholds. The value of $Q$ is obtained as the average value of $\mu_R = \hat{H}_\mathrm{T}$ calculated at NNLO for each $H_{\mathrm{T}2}$ bin.

The values of $\alpha_s(m_\Zboson)$ obtained from global fits to the TEEC and ATEEC distributions, using the \textsc{MMHT 2014}, \textsc{CT14} and \textsc{NNPDF 3.0} PDF sets are shown in Tables~\ref{tab:aspdfTEEC} and~\ref{tab:aspdfATEEC}, together with the corresponding $\chi^2$ values. The values of $\alpha_{\mathrm{s}}(m_\Zboson)$ determined from fits to the TEEC functions, both in the inclusive and exclusive $H_{\mathrm{T}2}$ bins, are presented in Table~\ref{tab:teecAlpha}, while the values obtained from fits to the ATEEC functions are presented in Table~\ref{tab:ateecAlpha}. The average $Q$ values and the reduced $\chi^2$ values at the minimum are also shown. The values displayed in Tables~\ref{tab:teecAlpha} and~\ref{tab:ateecAlpha} are obtained using the \textsc{MMHT 2014} PDF in the theoretical predictions.
 
\begin{table}[H]
\caption{Values of $\alpha_s(m_\Zboson)$ obtained from fits to the TEEC distributions to the NNLO theoretical predictions obtained using different PDF sets. The uncertainty referred to as $\mu$ represents the theoretical scale uncertainty. The uncertainty referred to as NP is related to the non-pQCD corrections. The uncertainty referred to as (mod.) is related to the MC model used in the unfolding. The value of the $\chi^2$ function at its minimum, as well as the number of degrees of freedom is also indicated.}
\label{tab:aspdfTEEC}
\centering
\footnotesize
\vspace{0.5cm}
\begin{tabular}{ c c c }
\hline
\T\B PDF & $\alpha_s(m_\Zboson)$ value & $\chi^2 / N_\mathrm{dof}$\\
\hline
\T MMHT 2014 & $0.1175 \pm 0.0001 \mbox{ (stat.)} \pm 0.0006 \mbox{ (sys.)}^{+0.0032}_{-0.0011} \mbox{ ($\mu$)} \pm 0.0011 \mbox{ (PDF)} \pm 0.0002 \mbox{ (NP)} \pm 0.0005 \mbox{ (mod.)}$ & 318 / 251\\
\T CT14 & $0.1196 \pm 0.0001 \mbox{ (stat.)} \pm 0.0006 \mbox{ (sys.)}^{+0.0035}_{-0.0010} \mbox{ ($\mu$)} \pm 0.0016 \mbox{ (PDF)} \pm 0.0002 \mbox{ (NP)} \pm 0.0006 \mbox{ (mod.)}$ & 262 / 251\\
\T\B NNPDF 3.0 & $0.1191 \pm 0.0001 \mbox{ (stat.)} \pm 0.0006 \mbox{ (sys.)}^{+0.0040}_{-0.0011} \mbox{ ($\mu$)} \pm 0.0020 \mbox{ (PDF)} \pm 0.0003 \mbox{ (NP)} \pm 0.0007 \mbox{(mod.)}$ & 300 / 251\\
\hline
\end{tabular}
\end{table}
 
\begin{table}[H]
\caption{Values of $\alpha_s(m_\Zboson)$ obtained from fits to the ATEEC distributions to the NNLO theoretical predictions obtained using different PDF sets. The uncertainty referred to as $\mu$ represents the theoretical scale uncertainty. The uncertainty referred to as NP is related to the non-pQCD corrections. The uncertainty referred to as (mod.) is related to the MC model used in the unfolding. The value of the $\chi^2$ function at its minimum, as well as the number of degrees of freedom is also indicated.}
\label{tab:aspdfATEEC}
\centering
\footnotesize
\vspace{0.5cm}
\begin{tabular}{ c c c }
\hline
\T\B PDF & $\alpha_s(m_\Zboson)$ value & $\chi^2 / N_\mathrm{dof}$\\
\hline
\T MMHT 2014 & $0.1185 \pm 0.0005 \mbox{ (stat.)} \pm 0.0008 \mbox{ (sys.)}^{+0.0022}_{-0.0002} \mbox{ ($\mu$)} \pm 0.0011 \mbox{ (PDF)} \pm 0.0004 \mbox{ (NP)} \pm 0.0001 \mbox{ (mod.)}$ & 110 / 117\\
\T CT14 & $0.1200 \pm 0.0006 \mbox{ (stat.)} \pm 0.0009 \mbox{ (sys.)}^{+0.0027}_{-0.0001} \mbox{ ($\mu$)} \pm 0.0016 \mbox{ (PDF)} \pm 0.0005 \mbox{ (NP)} \pm 0.0001 \mbox{ (mod.)}$ & 110 / 117\\
\T\B NNPDF 3.0 & $0.1199 \pm 0.0006 \pm \mbox{ (stat.)} 0.0009 \mbox{ (sys.)}^{+0.0027}_{-0.0002} \mbox{ ($\mu$)} \pm 0.0017 \mbox{ (PDF)} \pm 0.0005 \mbox{ (NP)} \pm 0.0001 \mbox{(mod.)}$ & 108 / 117\\
\hline
\end{tabular}
\end{table}
 
The central values of $\alpha_s$, obtained with different PDFs, in Tables~\ref{tab:aspdfTEEC} and~\ref{tab:aspdfATEEC} show differences that are, in some cases, larger than the uncertainty band on the eigenvectors or replicas provided for each PDF set. This is understood to arise from the differences between PDFs at high values of the Bjorken variable $x$, which influence the theory description of the TEEC and ATEEC in the phase space region relevant for this analysis.
 
Since \textsc{MMHT 2014} provides the largest range on $\alpha_s(m_Z)$ among the three PDF sets investigated, it is used as the central PDF set. The final values of $\alpha_s$ obtained from global fits to the TEEC and ATEEC distributions using the \textsc{MMHT 2014} PDF are
\[
\begin{array}{c}
\B\alpha_\mathrm{s}(m_\Zboson) = 0.1175 \pm 0.0006 \mbox{ (exp.)} ^{+0.0034}_{-0.0017} \mbox{ (theo.) and}\\
\T\alpha_\mathrm{s}(m_\Zboson) = 0.1185 \pm 0.0009 \mbox{ (exp.)} ^{+0.0025}_{-0.0012} \mbox{ (theo.)}.
\end{array}
\]
The central values of $\alpha_s$ determined from both observables are correlated with a Pearson correlation coefficient $\rho=0.86\pm 0.02 \mbox{ (exp.)}$, and are found to be compatible within the quoted uncertainties. While the value determined from the fit to the TEEC distributions has a better experimental precision, mainly due to its smaller statistical uncertainty, the value determined from the fit to the ATEEC distributions exhibits a better theoretical precision.
 
The fitted values of $\alpha_{\mathrm{s}}(m_\Zboson)$ for each bin, together with the values of the nuisance parameters, are used to evaluate the agreement between the data and the fitted predictions. Figure~\ref{fig:teecRatio} shows the ratio of the TEEC data to the fitted theoretical distributions for both the inclusive and exclusive bins, together with the theoretical and experimental uncertainties. Similarly, Figure~\ref{fig:ateecRatio} shows the ratios of data to theory for the ATEEC distributions. An excellent agreement is observed for both the TEEC and ATEEC measurements in all regions of phase space.
 
The values of $\alpha_{\mathrm{s}}(m_\Zboson)$ obtained in Tables~\ref{tab:teecAlpha} and~\ref{tab:ateecAlpha} are evolved to the corresponding energy scale using the RGE as explained above. This allows testing the asymptotic behaviour of QCD by comparing the measured points with the prediction from the RGE\@. Figures~\ref{fig:teecRunning} and~\ref{fig:ateecRunning} show the values of $\alpha_{\mathrm{s}}(Q)$ together with the world average band provided by the Particle Data Group~\cite{PDG2022} and values of $\alpha_\mathrm{s}$ obtained in other analyses~\cite{D0-1, D0-2, STDM-2014-03, STDM-2016-10, D0-1, STDM-2012-19, CMS-SMP-14-001, CMS-SMP-12-027, CMS-QCD-11-003, CMS-TOP-18-004}. To compare the results with other experiments, the value of $Q$ is chosen as half of the average value of $\hat{H}_\mathrm{T}$ for each $H_{\mathrm{T}2}$ bin. The results show a good agreement between all measurements and the RGE prediction. The values of $\alpha_s$ obtained from fits to the theoretical predictions using different PDF sets, shown in Figs.~\ref{fig:teecRunningPDF} and~\ref{fig:ateecRunningPDF}, are found to be in good agreement. Since the main purpose of the present analysis is to provide a first determination of $\alpha_s$ using NNLO predictions for three-jet cross sections, a combined determination of $\alpha_s$ and the PDF~\cite{forte} is not attempted. Such analysis would require to use very large computing resources and is well out of the scope of this paper.
 
\newpage
\begin{landscape}
\begin{table}[H]
\caption{Values of the strong coupling constant at the $Z$ boson mass scale, $\alpha_{\mathrm{s}}(m_\Zboson)$, obtained from fits to the TEEC function using \textsc{MMHT 2014} parton distribution functions. The value referred to as `inclusive' is obtained from a fit to the TEEC distribution for $H_{\mathrm{T}2} > 1~\TeV$. The values of the average interaction scale $\langle Q \rangle$ are shown in the first column, while the values of the $\chi^{2}$ function at the minimum are shown in the third column. The uncertainty referred to as $\mu$ represents the theoretical scale uncertainty. The uncertainty referred to as NP is related to the non-pQCD corrections. The uncertainty referred to as (mod.) is related to the MC model used in the unfolding.}
\label{tab:teecAlpha}
\small
\vspace{2.0cm}
\centering
\begin{tabular}{ c c c }
\hline
\T\B $\langle \hat{H}_{\mathrm{T}} \rangle$ [\GeV] & $\alpha_{\mathrm{s}}(m_\Zboson)$ value (\textsc{MMHT 2014}) & $\chi^{2} / N_{\mathrm{dof}}$ \\
\hline
\T\B Inclusive & 0.1188 $\pm$ 0.0002 (stat.) $\pm$ 0.0007 (syst.) $^{+0.0030}_{-0.0002}$ ($\mu$) $\pm$ 0.0010 (PDF) $\pm$ 0.0002 (NP) $\pm$ 0.0008 (mod.) & 16.0 / 27 \\
\hline
\T\B 1302 & 0.1186 $\pm$ 0.0003 (stat.) $\pm$ 0.0009 (syst.) $^{+0.0031}_{-0.0001}$ ($\mu$) $\pm$ 0.0010 (PDF) $\pm$ 0.0003 (NP) $\pm$ 0.0007 (mod.) & 16.4 / 27 \\
\T\B 1518 & 0.1182 $\pm$ 0.0003 (stat.) $\pm$ 0.0009 (syst.) $^{+0.0027}_{-0.0005}$ ($\mu$) $\pm$ 0.0010 (PDF) $\pm$ 0.0003 (NP) $\pm$ 0.0008 (mod.) & 15.1 / 27 \\
\T\B 1732 & 0.1191 $\pm$ 0.0003 (stat.) $\pm$ 0.0011 (syst.) $^{+0.0030}_{-0.0005}$ ($\mu$) $\pm$ 0.0011 (PDF) $\pm$ 0.0005 (NP) $\pm$ 0.0010 (mod.) & 19.8 / 27 \\
\T\B 1944 & 0.1179 $\pm$ 0.0003 (stat.) $\pm$ 0.0011 (syst.) $^{+0.0030}_{-0.0005}$ ($\mu$) $\pm$ 0.0011 (PDF) $\pm$ 0.0005 (NP) $\pm$ 0.0009 (mod.) & 21.3 / 27 \\
\T\B 2153 & 0.1175 $\pm$ 0.0004 (stat.) $\pm$ 0.0012 (syst.) $^{+0.0029}_{-0.0004}$ ($\mu$) $\pm$ 0.0011 (PDF) $\pm$ 0.0003 (NP) $\pm$ 0.0009 (mod.) & 32.9 / 27 \\
\T\B 2396 & 0.1179 $\pm$ 0.0003 (stat.) $\pm$ 0.0012 (syst.) $^{+0.0029}_{-0.0005}$ ($\mu$) $\pm$ 0.0012 (PDF) $\pm$ 0.0011 (NP) $\pm$ 0.0012 (mod.) & 27.5 / 27 \\
\T\B 2706 & 0.1164 $\pm$ 0.0004 (stat.) $\pm$ 0.0015 (syst.) $^{+0.0030}_{-0.0005}$ ($\mu$) $\pm$ 0.0013 (PDF) $\pm$ 0.0005 (NP) $\pm$ 0.0011 (mod.) & 35.1 / 27 \\
\T\B 3042 & 0.1162 $\pm$ 0.0005 (stat.) $\pm$ 0.0017 (syst.) $^{+0.0031}_{-0.0005}$ ($\mu$) $\pm$ 0.0013 (PDF) $\pm$ 0.0002 (NP) $\pm$ 0.0015 (mod.) & 33.1 / 27 \\
\T\B 3476 & 0.1141 $\pm$ 0.0007 (stat.) $\pm$ 0.0017 (syst.) $^{+0.0033}_{-0.0011}$ ($\mu$) $\pm$ 0.0014 (PDF) $\pm$ 0.0002 (NP) $\pm$ 0.0020 (mod.) & 15.1 / 13 \\
\T\B 4189 & 0.1116 $\pm$ 0.0011 (stat.) $\pm$ 0.0018 (syst.) $^{+0.0030}_{-0.0009}$ ($\mu$) $\pm$ 0.0015 (PDF) $\pm$ 0.0002 (NP) $\pm$ 0.0020 (mod.) & 14.0 / 13 \\
\hline
\end{tabular}
\end{table}
\end{landscape}
 
\begin{landscape}
\begin{table}[H]
\caption{Values of the strong coupling constant at the $Z$ boson mass scale, $\alpha_{\mathrm{s}}(m_\Zboson)$, obtained from fits to the ATEEC function using \textsc{MMHT 2014} parton distribution functions. The value referred to as `inclusive' is obtained from a fit to the ATEEC distribution for $H_{\mathrm{T}2} > 1~\TeV$. The values of the average interaction scale $\langle Q \rangle$ are shown in the first column, while the values of the $\chi^{2}$ function at the minimum are shown in the third column. The uncertainty referred to as $\mu$ represents the theoretical scale uncertainty. The uncertainty referred to as NP is related to the non-pQCD corrections. The uncertainty referred to as (mod.) is related to the MC model used in the unfolding.}
\label{tab:ateecAlpha}
\small
\vspace{2.0cm}
\centering
\begin{tabular}{ c c c }
\hline
\T\B $\langle \hat{H}_{\mathrm{T}} \rangle$ [\GeV] & $\alpha_{\mathrm{s}}(m_\Zboson)$ value (\textsc{MMHT 2014}) & $\chi^{2} / N_{\mathrm{dof}}$ \\
\hline
\T\B Inclusive & 0.1194 $\pm$ 0.0009 (stat.) $\pm$ 0.0007 (syst.) $^{+0.0023}_{-0.0000}$ ($\mu$) $\pm$ 0.0011 (PDF) $\pm$ 0.0005 (NP) $\pm$ 0.0000 (mod.) & 10.7 / 12 \\
\hline
\T\B 1302 & 0.1195 $\pm$ 0.0011 (stat.) $\pm$ 0.0006 (syst.) $^{+0.0024}_{-0.0000}$ ($\mu$) $\pm$ 0.0011 (PDF) $\pm$ 0.0006 (NP) $\pm$ 0.0000 (mod.) & 10.6 / 12 \\
\T\B 1518 & 0.1191 $\pm$ 0.0011 (stat.) $\pm$ 0.0007 (syst.) $^{+0.0020}_{-0.0001}$ ($\mu$) $\pm$ 0.0011 (PDF) $\pm$ 0.0004 (NP) $\pm$ 0.0001 (mod.) & 7.2 / 12 \\
\T\B 1732 & 0.1187 $\pm$ 0.0015 (stat.) $\pm$ 0.0009 (syst.) $^{+0.0026}_{-0.0003}$ ($\mu$) $\pm$ 0.0012 (PDF) $\pm$ 0.0010 (NP) $\pm$ 0.0002 (mod.) & 7.2 / 12 \\
\T\B 1944 & 0.1178 $\pm$ 0.0016 (stat.) $\pm$ 0.0009 (syst.) $^{+0.0022}_{-0.0001}$ ($\mu$) $\pm$ 0.0013 (PDF) $\pm$ 0.0007 (NP) $\pm$ 0.0000 (mod.) & 11.2 / 12 \\
\T\B 2153 & 0.1174 $\pm$ 0.0017 (stat.) $\pm$ 0.0009 (syst.) $^{+0.0022}_{-0.0002}$ ($\mu$) $\pm$ 0.0013 (PDF) $\pm$ 0.0007 (NP) $\pm$ 0.0001 (mod.) & 12.8 / 12 \\
\T\B 2396 & 0.1187 $\pm$ 0.0017 (stat.) $\pm$ 0.0010 (syst.) $^{+0.0017}_{-0.0000}$ ($\mu$) $\pm$ 0.0012 (PDF) $\pm$ 0.0007 (NP) $\pm$ 0.0004 (mod.) & 11.7 / 12 \\
\T\B 2706 & 0.1148 $\pm$ 0.0026 (stat.) $\pm$ 0.0014 (syst.) $^{+0.0024}_{-0.0001}$ ($\mu$) $\pm$ 0.0015 (PDF) $\pm$ 0.0007 (NP) $\pm$ 0.0002 (mod.) & 18.3 / 12 \\
\T\B 3042 & 0.1169 $\pm$ 0.0031 (stat.) $\pm$ 0.0012 (syst.) $^{+0.0018}_{-0.0000}$ ($\mu$) $\pm$ 0.0015 (PDF) $\pm$ 0.0014 (NP) $\pm$ 0.0009 (mod.) & 13.0 / 12 \\
\T\B 3476 & 0.1141 $\pm$ 0.0052 (stat.) $\pm$ 0.0016 (syst.) $^{+0.0025}_{-0.0007}$ ($\mu$) $\pm$ 0.0018 (PDF) $\pm$ 0.0008 (NP) $\pm$ 0.0011 (mod.) & 4.9 / 6 \\
\T\B 4189 & 0.1096 $\pm$ 0.0085 (stat.) $\pm$ 0.0009 (syst.) $^{+0.0013}_{-0.0000}$ ($\mu$) $\pm$ 0.0009 (PDF) $\pm$ 0.0002 (NP) $\pm$ 0.0007 (mod.) & 6.2 / 6 \\
\hline
\end{tabular}
\end{table}
\end{landscape}
 
\clearpage
\begin{figure}[H]
\vspace{2.cm}
\centering
\includegraphics[width=1.00\textwidth]{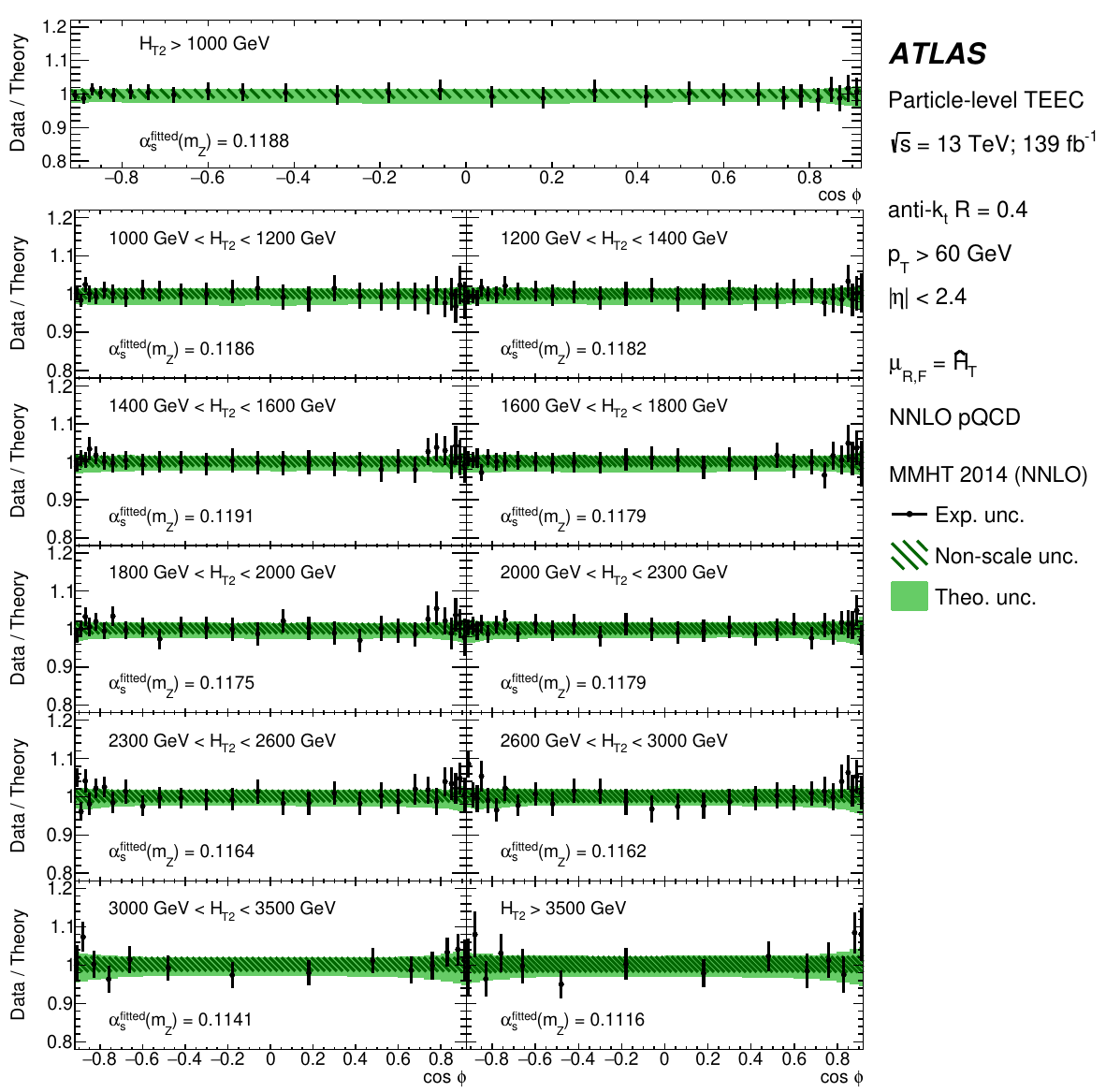}
\caption{Ratios of the data to the fitted theoretical predictions for the TEEC measurements in inclusive and exclusive $H_{\mathrm{T}2}$ bins. The solid band shows the theoretical uncertainties, dominated by the scale variations, while the error bars show the experimental uncertainties, where correlations between the fit parameters are taken into account. The hatched band shows the theoretical uncertainties, other than the scale variations. The central values of $\alpha_s(m_Z)$ determined for each $H_{\mathrm{T}2}$ bin are indicated.}
\label{fig:teecRatio}
\end{figure}

\clearpage
\begin{figure}[H]
\vspace{2.cm}
\centering
\includegraphics[width=1.00\textwidth]{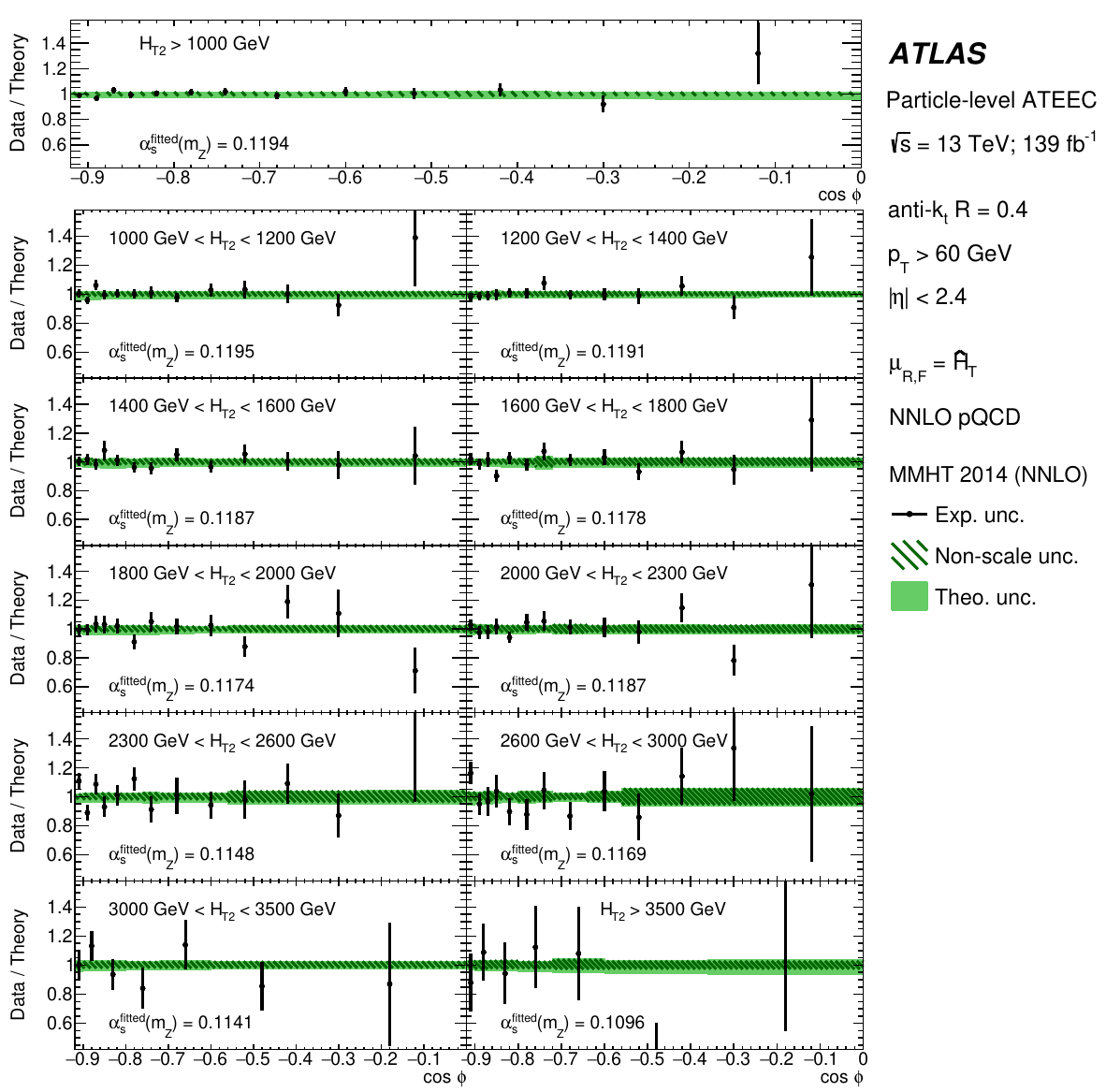}
\caption{Ratios of the data to the fitted theoretical predictions for the ATEEC measurements in inclusive and exclusive $H_{\mathrm{T}2}$ bins. The solid band shows the theoretical uncertainties, dominated by the scale variations, while the error bars show the experimental uncertainties, where correlations between the fit parameters are taken into account. The hatched band shows the theoretical uncertainties, other than the scale variations. The central values of $\alpha_s(m_Z)$ determined for each $H_{\mathrm{T}2}$ bin are indicated.}
\label{fig:ateecRatio}
\end{figure}
 
\clearpage
\begin{figure}[H]
\vspace{-0.3cm}
\centering
\includegraphics[width=0.83\textwidth]{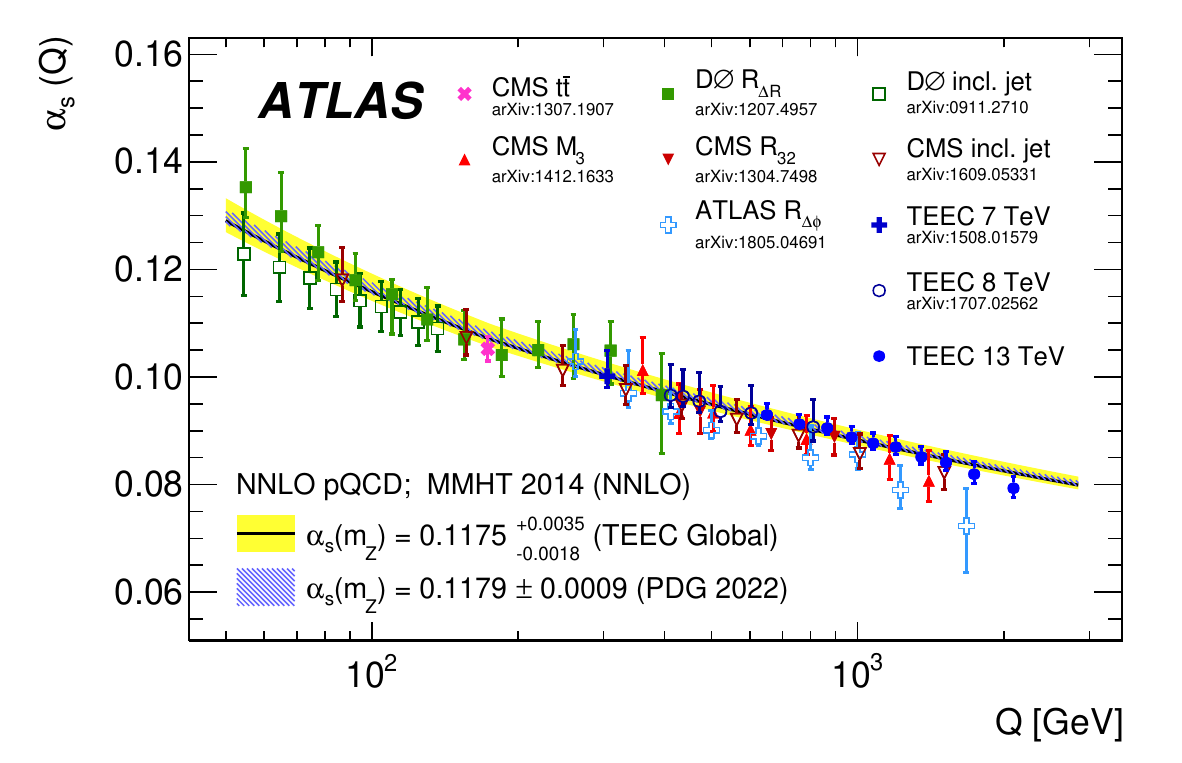}
\caption{Comparison of the values of $\alpha_\mathrm{s}(Q)$ determined from fits to the TEEC functions with the QCD prediction using the world average as input (hatched band) and the value obtained from the global fit (solid band). Results from previous analyses, both from ATLAS and from other experiments, are also included, showing an excellent agreement with the current measurements and with the world average. The value of $Q$ for the TEEC 13~\TeV\ points is chosen as half of the average of $\hat{H}_{\mathrm{T}}$ in each $H_{\mathrm{T}2}$ bin.}
\label{fig:teecRunning}
\end{figure}
 
\begin{figure}[H]
\vspace{-0.3cm}
\centering
\includegraphics[width=0.83\textwidth]{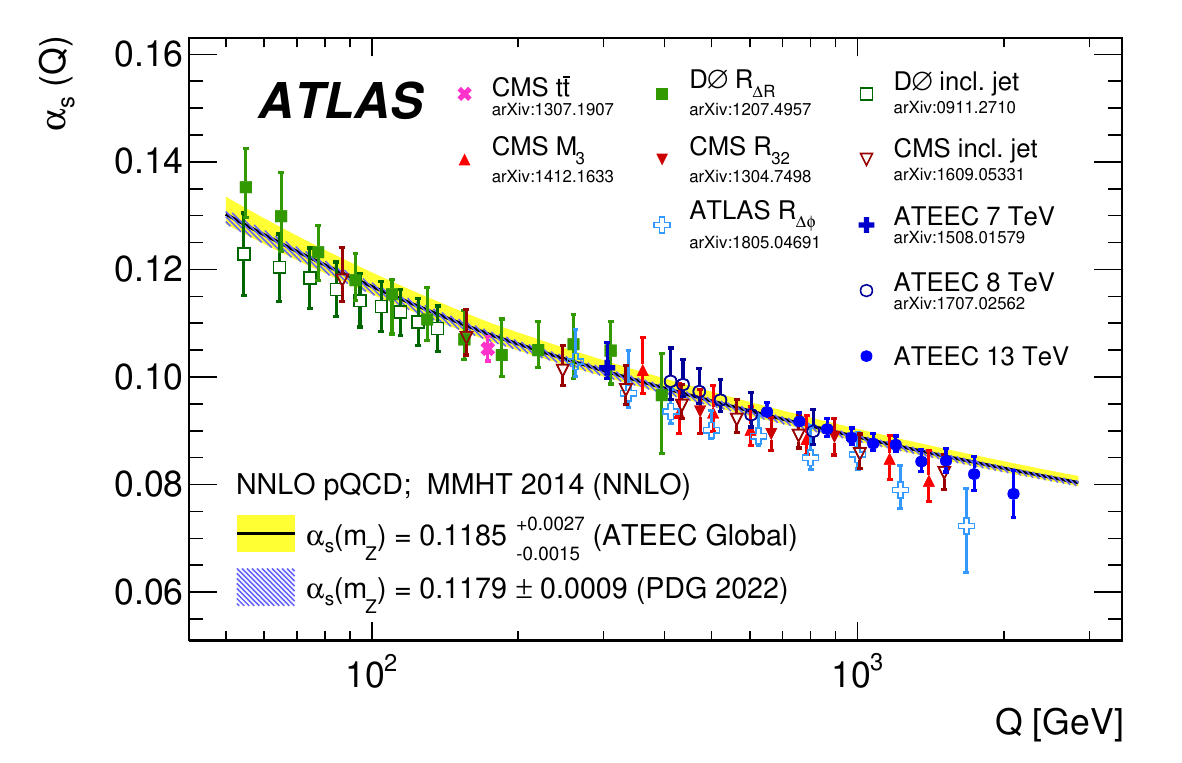}
\caption{Comparison of the values of $\alpha_\mathrm{s}(Q)$ determined from fits to the ATEEC functions with the QCD prediction using the world average as input (hatched band) and the value obtained from the global fit (solid band). Results from previous analyses, both from ATLAS and from other experiments, are also included, showing an excellent agreement with the current measurements and with the world average. The value of $Q$ for the TEEC 13~\TeV\ points is chosen as half of the average of $\hat{H}_{\mathrm{T}}$ in each $H_{\mathrm{T}2}$ bin.}
\label{fig:ateecRunning}
\end{figure}
 
\clearpage
 
\begin{figure}[H]
\vspace{0.7cm}
\centering
\includegraphics[width=0.83\textwidth]{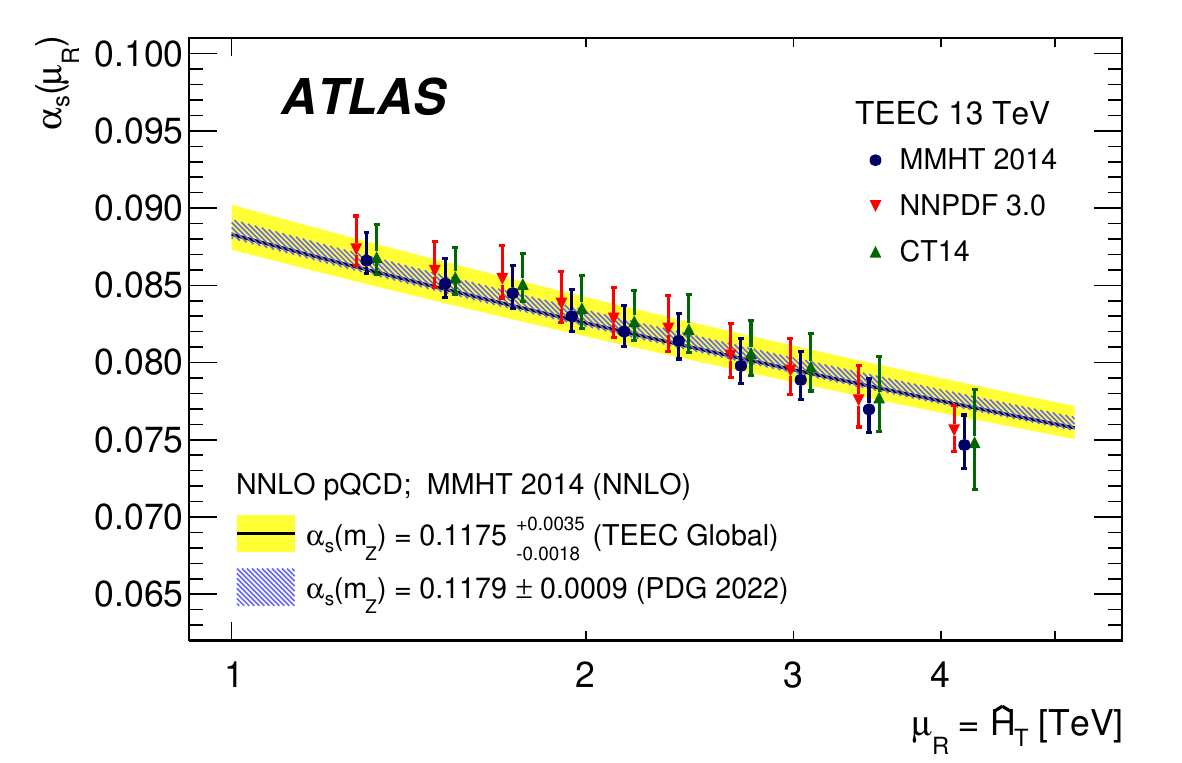}
\caption{Values of $\alpha_s(\mu_R)$ determined in each bin of $H_{\mathrm{T}2}$ from fits to the TEEC distributions using theoretical predictions with the \textsc{MMHT2014}, \textsc{CT14} and \textsc{NNPDF 3.0} PDF sets.}
\label{fig:teecRunningPDF}
\end{figure}
 
\begin{figure}[H]
\vspace{-0.3cm}
\centering
\includegraphics[width=0.83\textwidth]{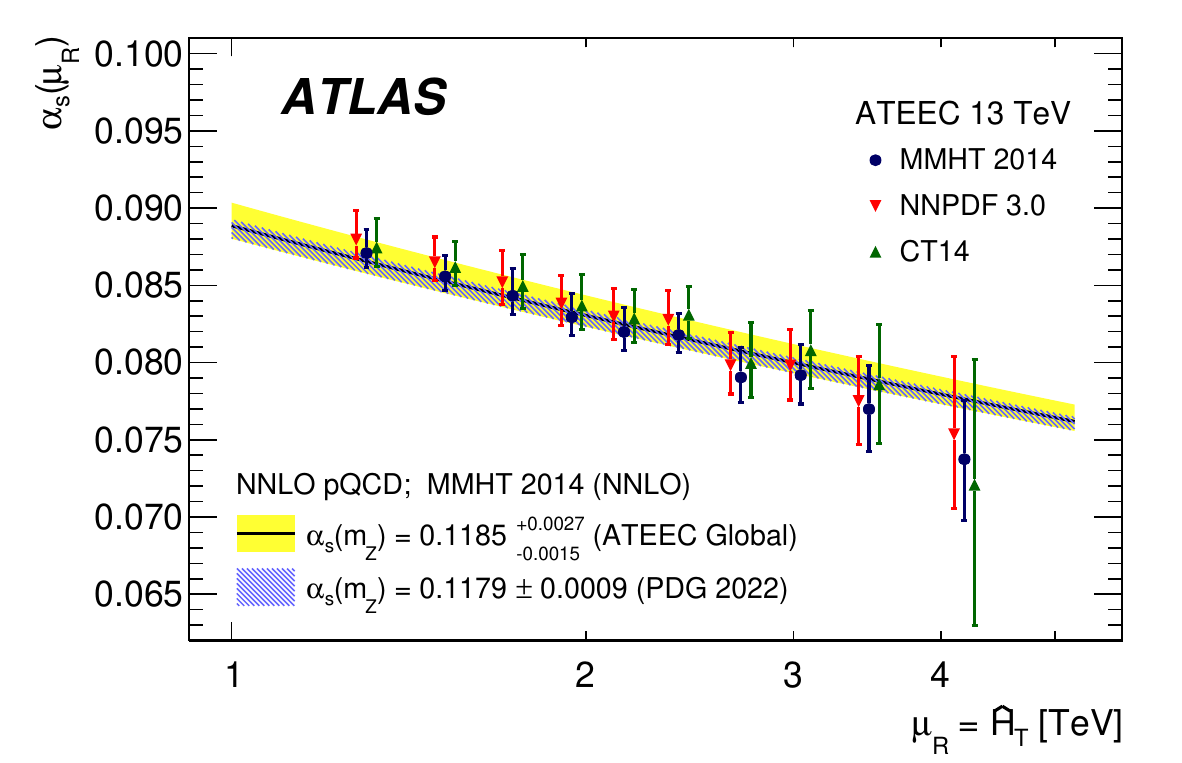}
\caption{Values of $\alpha_s(\mu_R)$ determined in each bin of $H_{\mathrm{T}2}$ from fits to the ATEEC distributions using theoretical predictions with the \textsc{MMHT2014}, \textsc{CT14} and \textsc{NNPDF 3.0} PDF sets.}
\label{fig:ateecRunningPDF}
\end{figure}


\FloatBarrier

\section{Conclusions}
A measurement of transverse energy--energy correlations and their corresponding asymmetries in multijet events using 139~\ifb of $pp$ collisions at $\sqrt{s} = 13~\TeV$ collected with the ATLAS detector at the LHC is presented. High-energy multijet events are selected by requiring the scalar sum of $\pt$ of the two leading jets, $H_{\mathrm{T}2}$, to be above 1~\TeV, and the data are binned in this variable to study the scale dependence of these observables.
 
The data are corrected for detector effects and systematic uncertainties are evaluated. The experimental uncertainties are dominated by the uncertainty in the jet energy scale and the model used to correct for detector effects. The total uncertainty is found to be of the order of 2\% for the TEEC and 1\% for the ATEEC\@. The results are compared with MC predictions from different generators, including \textsc{Pythia~8}, \textsc{Sherpa} and \textsc{Herwig 7} with two different parton showers, one angle-ordered and one based on dipole radiation.
 
The data are also compared with novel theoretical predictions at next-to-next-to-leading order in perturbative QCD, which are corrected for non-perturbative effects such as hadronisation and multiple parton interactions. The theoretical uncertainty due to the choice of the renormalisation and factorisation scales amount to 2\% for the TEEC and 3\% for the ATEEC, being the dominant uncertainty for both observables. The agreement between the data and the theoretical predictions is excellent. From this comparison, values of $\alpha_{\mathrm{s}}(m_\Zboson)$ are obtained for both the $H_{\mathrm{T}2}$-inclusive measurement and for the different exclusive bins in $H_{\mathrm{T}2}$ using a $\chi^2$ fit. The values of $\alpha_\mathrm{s}(m_\Zboson)$ obtained from the global fit to all $H_{\mathrm{T}2}$ bins simultaneously, are
\[
\begin{array}{c}
\B\alpha_\mathrm{s}(m_\Zboson) = 0.1175 \pm 0.0006 \mbox{ (exp.)} ^{+0.0034}_{-0.0017} \mbox{ (theo.) and}\\
\T\alpha_\mathrm{s}(m_\Zboson) = 0.1185 \pm 0.0009 \mbox{ (exp.)} ^{+0.0025}_{-0.0012} \mbox{ (theo.)}
\end{array}
\]
for the TEEC and ATEEC, respectively. The inclusion of next-to-next-to-leading-order corrections to three-jet production in the theoretical predictions for the first time has allowed a reduction by a factor of 3 of the theoretical uncertainties in both the cross-section calculation for the TEEC and ATEEC distributions and in the determination of the strong coupling constant $\alpha_s$.
 
The evolution of the values obtained from each of the exclusive fits to the corresponding measurement scales leads to values of $\alpha_\mathrm{s}(Q)$ that are compared with the next-to-next-to-leading-order solution of the RGE. The results show a good agreement with the RGE predictions up to the highest energy scales, and with previous measurements by ATLAS and by other experiments.


\section*{Acknowledgements}


We thank CERN for the very successful operation of the LHC, as well as the
support staff from our institutions without whom ATLAS could not be
operated efficiently.
 
We acknowledge the support of
ANPCyT, Argentina;
YerPhI, Armenia;
ARC, Australia;
BMWFW and FWF, Austria;
ANAS, Azerbaijan;
CNPq and FAPESP, Brazil;
NSERC, NRC and CFI, Canada;
CERN;
ANID, Chile;
CAS, MOST and NSFC, China;
Minciencias, Colombia;
MEYS CR, Czech Republic;
DNRF and DNSRC, Denmark;
IN2P3-CNRS and CEA-DRF/IRFU, France;
SRNSFG, Georgia;
BMBF, HGF and MPG, Germany;
GSRI, Greece;
RGC and Hong Kong SAR, China;
ISF and Benoziyo Center, Israel;
INFN, Italy;
MEXT and JSPS, Japan;
CNRST, Morocco;
NWO, Netherlands;
RCN, Norway;
MEiN, Poland;
FCT, Portugal;
MNE/IFA, Romania;
MESTD, Serbia;
MSSR, Slovakia;
ARRS and MIZ\v{S}, Slovenia;
DSI/NRF, South Africa;
MICINN, Spain;
SRC and Wallenberg Foundation, Sweden;
SERI, SNSF and Cantons of Bern and Geneva, Switzerland;
MOST, Taiwan;
TENMAK, T\"urkiye;
STFC, United Kingdom;
DOE and NSF, United States of America.
In addition, individual groups and members have received support from
BCKDF, CANARIE, Compute Canada and CRC, Canada;
PRIMUS 21/SCI/017 and UNCE SCI/013, Czech Republic;
COST, ERC, ERDF, Horizon 2020 and Marie Sk{\l}odowska-Curie Actions, European Union;
Investissements d'Avenir Labex, Investissements d'Avenir Idex and ANR, France;
DFG and AvH Foundation, Germany;
Herakleitos, Thales and Aristeia programmes co-financed by EU-ESF and the Greek NSRF, Greece;
BSF-NSF and MINERVA, Israel;
Norwegian Financial Mechanism 2014-2021, Norway;
NCN and NAWA, Poland;
La Caixa Banking Foundation, CERCA Programme Generalitat de Catalunya and PROMETEO and GenT Programmes Generalitat Valenciana, Spain;
G\"{o}ran Gustafssons Stiftelse, Sweden;
The Royal Society and Leverhulme Trust, United Kingdom.
 
The crucial computing support from all WLCG partners is acknowledged gratefully, in particular from CERN, the ATLAS Tier-1 facilities at TRIUMF (Canada), NDGF (Denmark, Norway, Sweden), CC-IN2P3 (France), KIT/GridKA (Germany), INFN-CNAF (Italy), NL-T1 (Netherlands), PIC (Spain), ASGC (Taiwan), RAL (UK) and BNL (USA), the Tier-2 facilities worldwide and large non-WLCG resource providers. Major contributors of computing resources are listed in Ref.~\cite{ATL-SOFT-PUB-2021-003}.


\clearpage
\printbibliography

@article{evShapes1,
   author="{A. Banfi, G.P. Salam and G. Zanderighi}",
   title="{Resummed event shapes at hadron-hadron colliders}",
   journal="JHEP",
   volume="08",
   year="2004",
   pages="062",
   archivePrefix="arXiv",
   eprint="hep-ph/0407287",
   doi="10.1088/1126-6708/2004/08/062"
}

@article{evShapes2,
   author="{A. Banfi, G.P. Salam and G. Zanderighi}",
   title="{Phenomenology of event shapes at hadron colliders}",
   journal="JHEP",
   volume="06",
   year="2010",
   pages="038",
   archivePrefix="arXiv",
   eprint="1001.4082",
   primaryClass="hep-ph",
   doi="10.1007/JHEP06(2010)038"
}

@article{Aaltonen,
   author="{CDF Collaboration}",
   journal="Phys. Rev. D",
   volume="83",
   pages="112007",
   year=2011,
   archivePrefix="arXiv",
   eprint="1103.5143",
   primaryClass="hep-ex",
   doi="10.1103/PhysRevD.83.112007",
   title="{Measurement of event shapes in $p\overline{p}$ collisions at center-of-mass energy $1.96~\TeV$}"
}

@article{plutoEventShapes,
   author="{PLUTO Collaboration}",
   journal="Z. Phys. C",
   volume = "12",
   pages = "297",
   year=1982,
   doi="10.1007/BF01557575",
   title="{Energy dependence of jet measures in $e^+e^-$ annihilation}"
}

@article{tassoEventShapes,
   author="{TASSO Collaboration}",
   journal="Z. Phys. C",
   volume = "47",
   pages = "187",
   year=1990,
   doi="10.1007/BF01552339",
   title="{Global jet properties at 14-44 GeV center of mass energy in $e^+e^-$ annihilation}"
}

@article{jadeEventShapes,
   author="{JADE Collaboration}",
   journal="Eur. Phys. J. C",
   volume = "64",
   pages = "351",
   year=2009,
   archivePrefix="arXiv",
   eprint="0810.1389",
   primaryClass="hep-ex",
   doi="10.1140/epjc/s10052-009-1149-1",
   title = "{Determination of the strong coupling $\alpha_s$ from hadronic event shapes with $\mathcal{O}(\alpha_s^3)$ and resummed QCD predictions using JADE data}"
}

@article{thrustDelphi,
   author="{DELPHI Collaboration}",
   journal="Eur. Phys. J. C",
   volume = "29",
   pages = "285",
   year=2003,
   archivePrefix="arXiv",
   eprint="hep-ex/0307048",
   doi="10.1140/epjc/s2003-01198-0",
   title="{A study of the energy evolution of event shape distributions and their means with the DELPHI detector at LEP}"
}

@article{l3eventShapes,
   author="{L3 Collaboration}",
   journal="Phys. Lett. B",
   volume = "536",
   pages = "217",
   year=2002,
   archivePrefix="arXiv",
   eprint="hep-ex/0206052",
   doi="10.1016/S0370-2693(02)01814-2",
   title="{Determination of $\alpha_s$ from hadronic event shapes in $e^+e^-$ annihilation at $192 \leq \sqrt{s} \leq 208$ GeV}"
}

@article{opalEventShapes,
   author="{OPAL Collaboration}",
   journal="Eur. Phys. J. C",
   volume = "71",
   pages = "1733",
   year=2011,
   archivePrefix="arXiv",
   eprint="1101.1470",
   primaryClass="hep-ex",
   doi="10.1140/epjc/s10052-011-1733-z",
   title="{Determination of $\alpha_s$ using OPAL hadronic event shapes at $\sqrt{s} =$ 91 - 209 GeV and resummed NNLO calculations}"
}

@article{alephEventShapes,
   author="{ALEPH Collaboration}",
   journal="Eur. Phys. J. C",
   volume = "35",
   pages = "457",
   year=2003,
   doi="10.1140/epjc/s2004-01891-4",
   title="{Studies of QCD at $e^+e^-$ centre-of-mass energies between 91 and 209 GeV}"
}

@article{zeusEventShapes1,
   author="{ZEUS Collaboration}",
   journal="Eur. Phys. J. C",
   volume = "27",
   pages = "531",
   year=2003,
   archivePrefix="arXiv",
   eprint="hep-ex/0211040",
   doi="10.1140/epjc/s2003-01148-x",
   title="{Measurement of event shapes in deep inelastic scattering at HERA}"
}

@article{zeusEventShapes2,
   author="{ZEUS Collaboration}",
   journal="Nucl. Phys. B",
   volume = "767",
   pages = "1",
   year=2007,
   archivePrefix="arXiv",
   eprint="hep-ex/0604032",
   doi="10.1016/j.nuclphysb.2006.05.016",
   title="{Event shapes in deep inelastic scattering at HERA}"
}

@article{h1EventShapes1,
   author="{H1 Collaboration}",
   journal="Phys. Lett. B",
   volume = "406",
   pages = "256",
   year="1997",
   archivePrefix="arXiv",
   eprint="hep-ex/9706002",
   doi="10.1016/S0370-2693(97)00754-5",
   title="{Measurement of event shape variables in deep inelastic $ep$ scattering}"
}

@article{h1EventShapes2,
   author="{H1 Collaboration}",
   journal="Eur. Phys. J. C",
   volume = "14",
   pages = "255",
   year=2000,
   archivePrefix="arXiv",
   eprint="hep-ex/9912052",
   doi="10.1007/s100520000344",
   title="{Investigation of power corrections to event shape variables measured in deep-inelastic scattering}"
}

@article{h1EventShapes3,
   author="{H1 Collaboration}",
   journal="Eur. Phys. J. C",
   volume = "46",
   pages = "343",
   year=2006,
   archivePrefix="arXiv",
   eprint="hep-ex/0512014",
   doi="10.1140/epjc/s2006-02493-x",
   title="{Measurement of event shape variables in deep-inelastic scattering at HERA}"
}

@article{Khachatryan,
   author="{CMS Collaboration}",
   journal="Phys. Lett. B",
   volume = "699",
   pages = "48",
   year=2011,
   archivePrefix="arXiv",
   eprint="1102.0068",
   primaryClass="hep-ex",
   doi="10.1016/j.physletb.2011.03.060",
   title="{First measurement of hadronic event shapes in $pp$ collisions at $\sqrt{s}=7 ~\TeV$}"
}

@article{mitov2022,
    title={Next-to-Next-to-Leading Order Study of Three-Jet Production at the LHC},
    author={Michal Czakon and Alexander Mitov and Rene Poncelet},
    journal={Phys. Rev. Lett.},
    volume={127},
    year={2021},
    pages={152001},
    doi={10.1103/PhysRevLett.127.152001},
    eprint={2106.05331},
    archivePrefix={arXiv},
    primaryClass={hep-ph}
}

@article{mitov2023,
    title={NNLO QCD corrections to event shapes at the LHC},
    author={Manuel Álvarez and Josu Cantero and Michal Czakon and Javier Llorente and Alexander Mitov and Rene Poncelet},
    reportNumber="{Cavendish-HEP-22/11, P3H-22-129, TTK-22-49}",
    year="2023",
    eprint={2301.01086},
    archivePrefix={arXiv},
    primaryClass={hep-ph}
}

@Article{fivepoint,
    title = {Leading-color two-loop QCD corrections for three-jet production at hadron colliders},
    author={S. Abreu and F. Febres Cordero and H. Ita and B. Page and V. Sotnikov},
    journal={JHEP},
    volume={07},
    pages={95},
    year={2021},
    doi={10.1007/JHEP07(2021)095},
    eprint={2102.13609},
    archivePrefix={arXiv},
    primaryClass={hep-ph}
}

@Article{pentagonfunctions,
    title = {Pentagon functions for scattering of five massless particles},
    author={D. Chicherin and V. Sotnikov},
    journal={JHEP},
    volume={12},
    pages={167},
    year={2020},
    doi={10.1007/JHEP12(2020)167},
    eprint={2009.07803},
    archivePrefix={arXiv},
    primaryClass={hep-ph}
}

@Article{NNLOscheme1,
    title={Single-jet inclusive rates with exact color at $\mathcal{O}(\alpha_{\mathrm{s}})^{4}$},
    author={Michal Czakon and Andreas van Hameren and Alexander Mitov and Rene Poncelet},
    journal={JHEP},
    volume={10},
    pages={262},
    year={2019},
    doi={10.1007/JHEP10(2019)262},
    eprint={1907.12911},
    archivePrefix={arXiv},
    primaryClass={hep-ph}
}

@Article{NNLOscheme2,
    title={Four-dimensional formulation of the sector-improved residue subtraction scheme},
    author={Michal Czakon and David Heymes},
    journal={Nucl. Phys. B},
    volume={890},
    year={2015},
    pages={152},
    doi={10.1016/j.nuclphysb.2014.11.006},
    eprint={1408.2500},
    archivePrefix={arXiv},
    primaryClass={hep-ph}
}

@article{Aad,
   author="{ATLAS Collaboration}",
   journal="Eur. Phys. J. C",
   volume = "72",
   pages="2211",
   year=2012,
   archivePrefix="arXiv",
   eprint="1206.2135",
   primaryClass="hep-ex",
   doi="10.1140/epjc/s10052-012-2211-y",
   title="{Measurement of event shapes at large momentum transfer with the ATLAS detector in $pp$ collisions at $\sqrt{s} = 7 ~\TeV$}"
}

@Article{evShapes,
  author = "{ATLAS Collaboration}",
  journal = "JHEP",
  volume = "01",
  pages = "188",
  year = "2021",
  archivePrefix = "arXiv",
  eprint = "2007.12600",
  primaryClass = "hep-ex",
  doi = "10.1007/JHEP01(2021)188",
  title = "{Measurement of hadronic event shapes in high-pT multijet final states at $\sqrt{s} = 13$ TeV with the ATLAS detector}"
}

@Article{DESY-88-081,
  author = "{N.K. Falck and G. Kramer}",
  journal = "Z. Phys. C",
  volume = "42",
  year = 1989,
  pages = 459,
  doi = "10.1007/BF01548452",
  title = "{Theoretical studies of energy-energy correlation in $e^+ e^-$ annihilation}"
}

@article{sherpa,
   author="{T. Gleisberg \emph{et al.}}",
   title="{Event generation with SHERPA 1.1}",
   journal="JHEP",
   volume="02",
   year="2009",
   pages="007",
   archivePrefix="arXiv",
   eprint="0811.4622",
   doi="10.1088/1126-6708/2009/02/007"
}

@Article{sherpa22,
	title={{Event generation with Sherpa 2.2}},
	author="{E. Bothmann \emph{et al.}}",
	journal={SciPost Phys.},
	volume={7},
	pages={034},
	year={2019},
	publisher={SciPost},
	doi={10.21468/SciPostPhys.7.3.034},
	url={https://scipost.org/10.21468/SciPostPhys.7.3.034},
}

@article{CMWscale,
  author = "{S. Catani, B. R. Webber and G. Marchesini}",
  title = "{QCD coherent branching and semi-inclusive processes at large $\chi$}",
  journal = "Nucl. Phys. B",
  volume = "349",
  year = "1991",
  pages = "635-654",
  doi = "10.1016/0550-3213(91)90390-J",
}

@article{Krauss:2002up,
  author = "{F. Krauss}",
  title = "{Matrix Elements and Parton Showers in Hadronic Interactions}",
  journal = "JHEP",
  volume = "08",
  year = "2002",
  pages = "015",
  doi = "10.1088/1126-6708/2002/08/015",
  eprint = "hep-ph/0205283",
  archivePrefix = "arXiv",
  primaryClass = "hep-ph"
}

@article{mandelstam,
  author = "{S. Mandelstam}",
  title = "{Determination of the Pion-Nucleon Scattering Amplitude from Dispersion Relations and Unitarity. General Theory}",
  journal = "Phys. Rev.",
  volume = "112",
  pages = "1344",
  year = "1958",
  doi = "10.1103/PhysRev.112.1344"
}

@article{Webber:1983if,
  author = "Webber, B. R.",
  title  = "{A QCD model for jet fragmentation including soft gluon interference}",
  journal = "Nucl. Phys. B",
  volume         = "238",
  year           = "1984",
  pages          = "492-528",
  doi            = "10.1016/0550-3213(84)90333-X",
  reportNumber   = "CERN-TH-3713",
  SLACcitation   = "%%CITATION = NUPHA,B238,492;%%"
}

@article{ct14,
   author="{S. Dulat \emph{et al.}}",
   title="{New parton distribution functions from a global analysis of quantum chromodynamics}",
   journal="Phys. Rev. D",
   volume="93",
   year="2016",
   pages="033006",
   archivePrefix="arXiv",
   eprint="1506.07443",
   primaryClass="hep-ph",
   doi="10.1103/PhysRevD.93.033006"
}

@article{ct10,
   author="{H.L. Lai \emph{et al.}}",
   title="{New parton distributions for collider physics}",
   journal="Phys. Rev. D",
   volume="82",
   year="2010",
   pages="074024",
   archivePrefix="arXiv",
   eprint="1007.2241",
   primaryClass="hep-ph",
   doi="10.1103/PhysRevD.82.074024",
}

@article{nnpdf3,
   author="{R.D. Ball \emph{et al.}}",
   title="{Parton distributions for the LHC run II}",
   journal="JHEP",
   volume="04",
   year=2015,
   pages="040",
   archivePrefix="arXiv",
   eprint="1410.8849",
   primaryClass="hep-ph",
   doi="10.1007/JHEP04(2015)040"
}

@article{Bellm:2017bvx,
      author         = "{J. Bellm \emph{et al.}}",
      title          = "{Herwig 7.1 Release Note}",
      year           = "2017",
      eprint         = "1705.06919",
      archivePrefix  = "arXiv",
      primaryClass   = "hep-ph",
      reportNumber   = "CERN-PH-TH-2017-109, CERN-TH-2017-109, MAN-HEP-2017-08,
                        UWTHPH-2017-10, IFJPAN-IV-2017-7, NIKHEF-2017-026,
                        HERWIG-2017-02, KA-TP-19-2017, MCNET-17-08, IPPP-17-40"
}

@article{Platzer:2011bc,
      author         = "{S. Plätzer and S. Gieseke}",
      title          = "{Dipole showers and automated NLO matching in Herwig++}",
      journal        = "Eur. Phys. J. C",
      volume         = "72",
      year           = "2012",
      pages          = "2187",
      doi            = "10.1140/epjc/s10052-012-2187-7",
      eprint         = "1109.6256",
      archivePrefix  = "arXiv",
      primaryClass   = "hep-ph",
      reportNumber   = "DESY-11-162, KA-TP-24-2011, HERWIG-11-01, MCNET-11-24",
}

@article{mmht,
     author="{L.A. Harland-Lang, A.D. Martin, P. Motylinski and R.S. Thorne}",
     title="{Parton distributions in the LHC era: MMHT 2014 PDFs}",
     journal="Eur. Phys. J. C",
     volume="75",
     year="2015",
     pages="204",
     archivePrefix="arXiv",
     eprint="1412.3989",
     primaryClass="hep-ph",
     doi="10.1140/epjc/s10052-015-3397-6"
}

@article{nnpdf,
    author="{R.D. Ball \emph{et al.}}",
    title="{Parton distributions with LHC data}",
    journal="Nucl. Phys. B",
    volume ="867",
    year="2013",
    pages="244",
    archivePrefix="arXiv",
    eprint="1207.1303",
    primaryClass="hep-ph",
    doi="10.1016/j.nuclphysb.2012.10.003"
}

@article{antikt,
    author="{M. Cacciari, G.P. Salam and G. Soyez}",
    title="{The anti-$k_t$ jet clustering algorithm}",
    journal="JHEP",
    volume="04",
    pages="063",
    year=2008,
    archivePrefix="arXiv",
    eprint="0802.1189",
    primaryClass="hep-ph",
    doi="10.1088/1126-6708/2008/04/063"
}

@article{pflow,
  author = "{ATLAS Collaboration}",
  title = "{Jet reconstruction and performance using particle flow with the ATLAS Detector}",
  journal = "Eur. Phys. J. C",
  volume = "77",
  year = "2017",
  pages = "466",
  archivePrefix = "arXiv",
  eprint="1703.10485",
  primaryClass="hep-ex",
  doi="10.1140/epjc/s10052-017-5031-2"
}

@article{JES1,
  author = "{ATLAS Collaboration}",
  title = "{Jet energy scale measurements and their systematic uncertainties in proton-proton collisions at $\sqrt{s} = 13~\TeV$ with the ATLAS detector.}",
  journal = "Phys. Rev. D",
  volume = "96",
  pages = "072002",
  year = "2017",
  archivePrefix="arXiv",
  eprint = "1703.09665",
  primaryClass = "hep-ex",
  doi="10.1103/PhysRevD.96.072002"
}

@article{JES2,
  author = "{ATLAS Collaboration}",
  title = "{Jet energy scale and resolution measured in proton-proton collisions at $\sqrt{s} = 13$ TeV with the ATLAS detector}",
  journal = "Eur. Phys. J. C",
  volume = "81",
  pages = "689",
  year = "2021",
  archivePrefix="arXiv",
  eprint = "2007.02645",
  primaryClass = "hep-ex",
  doi="10.1140/epjc/s10052-021-09402-3"
}

@Article{bayes,
    author="{G. D'Agostini}",
    title="{A multidimensional unfolding method based on Bayes' theorem}",
    journal="Nucl. Instrum. Meth. A \textbf{362}, 487",
    year=1995,
    doi="10.1016/0168-9002(95)00274-X"
}

@article{forte,
  author="{S. Forte and Z. Kassabov}",
  title="{Why $\alpha_s$ cannot be determined from hadronic processes without simultaneously determining the parton distributions}",
  journal = "Eur. Phys. J. C",
  volume = "80",
  pages = "182",
  year = "2020",
  archivePrefix="arXiv",
  eprint="2001.04986",
  primaryClass="hep-ph",
  doi = "10.1140/epjc/s10052-020-7748-6"
}

@Article{Fastjet,
      author         = "Cacciari, Matteo and Salam, Gavin P. and Soyez, Gregory",
      title          = "{FastJet user manual}",
      journal        = "Eur. Phys. J. C",
      volume         = "72",
      year           = "2012",
      pages          = "1896",
      doi            = "10.1140/epjc/s10052-012-1896-2",
      eprint         = "1111.6097",
      archivePrefix  = "arXiv",
      primaryClass   = "hep-ph",
      reportNumber   = "CERN-PH-TH-2011-297",
      SLACcitation   = "%%CITATION = ARXIV:1111.6097;%%"
}

@Article{Sjostrand:2006za,
     author    = "Sj{\"o}strand, Torbjorn and Mrenna, Stephen and Skands, Peter Z.",
     title     = "{PYTHIA 6.4 physics and manual}",
     journal   = "JHEP",
     volume    = "05",
     year      = "2006",
     pages     = "026",
     eprint    = "hep-ph/0603175",
     archivePrefix = "arXiv",
     doi       = "10.1088/1126-6708/2006/05/026",
     SLACcitation  = "%%CITATION = HEP-PH/0603175;%%"
}

@article{Skands:2014pea,
      author         = "Skands, Peter and Carrazza, Stefano and Rojo, Juan",
      title          = "{Tuning PYTHIA 8.1: the Monash 2013 Tune}",
      journal        = "Eur. Phys. J. C",
      volume         = "74",
      year           = "2014",
      number         = "8",
      pages          = "3024",
      doi            = "10.1140/epjc/s10052-014-3024-y",
      eprint         = "1404.5630",
      archivePrefix  = "arXiv",
      primaryClass   = "hep-ph",
      reportNumber   = "CERN-PH-TH-2014-069, MCNET-14-08, OUTP-14-05P",
      SLACcitation   = "%%CITATION = ARXIV:1404.5630;%%"
}

@article{Sjostrand:2014zea,
      author         = "Sj{\"o}strand, Torbj{\"o}rn and Ask, Stefan and Christiansen,
                        Jesper R. and Corke, Richard and Desai, Nishita and Ilten,
                        Philip and Mrenna, Stephen and Prestel, Stefan and
                        Rasmussen, Christine O. and Skands, Peter Z.",
      title          = "{An introduction to PYTHIA 8.2}",
      journal        = "Comput. Phys. Commun.",
      volume         = "191",
      year           = "2015",
      pages          = "159",
      doi            = "10.1016/j.cpc.2015.01.024",
      eprint         = "1410.3012",
      archivePrefix  = "arXiv",
      primaryClass   = "hep-ph",
      reportNumber   = "LU-TP-14-36, MCNET-14-22, CERN-PH-TH-2014-190,
                        FERMILAB-PUB-14-316-CD, DESY-14-178, SLAC-PUB-16122,
                        --FERMILAB-PUB-14-316-CD",
      SLACcitation   = "%%CITATION = ARXIV:1410.3012;%%"
}

@article{Sjostrand:1984ic,
      author         = "Sj{\"o}strand, Torbjorn",
      title          = "{Jet fragmentation of multiparton
                  configurations in a string framework}",
      journal        = "Nucl. Phys. B",
      volume         = "248",
      year           = "1984",
      pages          = "469-502",
      doi            = "10.1016/0550-3213(84)90607-2",
      reportNumber   = "DESY-T-84-01",
      SLACcitation   = "%%CITATION = NUPHA,B248,469;%%"                                                                                                       
}

@article{Andersson:1983ia,
      author         = "Andersson, Bo and Gustafson, G. and Ingelman, G. and
                        Sj{\"o}strand, T.",
      title          = "{Parton fragmentation and string dynamics}",
      journal        = "Phys. Rept.",
      volume         = "97",
      year           = "1983",
      pages          = "31-145",
      doi            = "10.1016/0370-1573(83)90080-7",
      reportNumber   = "LU-TP-83-10",
      SLACcitation   = "%%CITATION = PRPLC,97,31;%%"
}

@Article{Frixione:2002ik,
     author    = "Frixione, Stefano and Webber, Bryan R.",
     title     = "{Matching NLO QCD computations and parton shower simulations}",
     journal   = "JHEP",
     volume    = "06",
     year      = "2002",
     pages     = "029",
     eprint    = "hep-ph/0204244",
     archivePrefix = "arXiv"
}

@Article{Bahr:2008pv,
      author         = "B{\"a}hr, M. and others",
      title          = "{Herwig++ physics and manual}",
      journal        = "Eur. Phys. J. C",
      volume         = "58",
      year           = "2008",
      pages          = "639",
      doi            = "10.1140/epjc/s10052-008-0798-9",
      eprint         = "0803.0883",
      archivePrefix  = "arXiv",
      primaryClass   = "hep-ph",
      reportNumber   = "CERN-PH-TH-2008-038, CAVENDISH-HEP-08-03, KA-TP-05-2008,
                        DCPT-08-22, IPPP-08-11, CP3-08-05",
      SLACcitation   = "%%CITATION = ARXIV:0803.0883;%%"
}

@Article{Bellm:2015jjp,
      author         = "Bellm, Johannes and others",
      title          = "{Herwig 7.0/Herwig++ 3.0 release note}",
      journal        = "Eur. Phys. J. C",
      volume         = "76",
      year           = "2016",
      number         = "4",
      pages          = "196",
      doi            = "10.1140/epjc/s10052-016-4018-8",
      eprint         = "1512.01178",
      archivePrefix  = "arXiv",
      primaryClass   = "hep-ph",
      reportNumber   = "CERN-PH-TH-2015-289, MAN-HEP-2015-15, IFJPAN-IV-2015-13,
                        HERWIG-2015-01, KA-TP-18-2015, DCPT-15-142, MCNET-15-28,
                        IPPP-15-71, --HERWIG-2015-01",
      SLACcitation   = "%%CITATION = ARXIV:1512.01178;%%"
}

@article{Hoeche:2012yf,
      author         = "H{\"o}che, Stefan and Krauss, Frank and Sch{\"o}nherr, Marek and
                        Siegert, Frank",
      title          = "{QCD matrix elements + parton showers. The NLO case}",
      journal        = "JHEP",
      volume         = "04",
      year           = "2013",
      pages          = "027",
      doi            = "10.1007/JHEP04(2013)027",
      eprint         = "1207.5030",
      archivePrefix  = "arXiv",
      primaryClass   = "hep-ph",
      reportNumber   = "SLAC-PUB-15191, IPPP-12-52, DCPT-12-104, LPN12-081,
                        FR-PHENO-2012-017, MCNET-12-09, --FR-PHENO-2012-017",
      SLACcitation   = "%%CITATION = ARXIV:1207.5030;%%"
}

@article{Winter:2003tt,
      author         = "Winter, Jan-Christopher and Krauss, Frank and Soff,
                        Gerhard",
      title          = "{A modified cluster-hadronization model}",
      journal        = "Eur. Phys. J. C",
      volume         = "36",
      year           = "2004",
      pages          = "381-395",
      doi            = "10.1140/epjc/s2004-01960-8",
      eprint         = "hep-ph/0311085",
      archivePrefix  = "arXiv",
      XprimaryClass   = "hep-ph",
      reportNumber   = "CERN-TH-2003-272",
      SLACcitation   = "%%CITATION = HEP-PH/0311085;%%"
}

@article{Catani:2001cc,
      author         = "Catani, S. and Krauss, F. and Webber, B. R. and Kuhn, R.",
      title          = "{QCD Matrix Elements + Parton Showers}",
      journal        = "JHEP",
      volume         = "11",
      year           = "2001",
      pages          = "063",
      doi            = "10.1088/1126-6708/2001/11/063",
      eprint         = "hep-ph/0109231",
      archivePrefix  = "arXiv",
      reportNumber   = "CERN-TH-2000-367, CAVENDISH-HEP-00-03",
      SLACcitation   = "%%CITATION = HEP-PH/0109231;%%"
}

@article{Buccioni:2019sur,
    author = "Buccioni, Federico and Lang, Jean-Nicolas and Lindert, Jonas M. and Maierh{\"o}fer, Philipp and Pozzorini, Stefano and Zhang, Hantian and Zoller, Max F.",
    title = "{OpenLoops 2}",
    eprint = "1907.13071",
    archivePrefix = "arXiv",
    primaryClass = "hep-ph",
    reportNumber = "IPPP/19/62, FR-PHENO-2019-12, PSI-PR-19-15, ZU-TH 37/19",
    doi = "10.1140/epjc/s10052-019-7306-2",
    journal = "Eur. Phys. J. C",
    volume = "79",
    number = "10",
    pages = "866",
    year = "2019"
}

@article{Agostinelli:2002hh,
      author         = "{GEANT4 Collaboration} and Agostinelli, S. and others",
      title          = "{\textsc{Geant4} -- a simulation toolkit}",
      journal        = "Nucl. Instrum. Meth. A",
      volume         = "506",
      year           = "2003",
      pages          = "250",
      doi            = "10.1016/S0168-9002(03)01368-8",
      reportNumber   = "SLAC-PUB-9350, FERMILAB-PUB-03-339",
      SLACcitation   = "%%CITATION = NUIMA,A506,250;%%"
}

@Article{PERF-2007-01,
    author         = "{ATLAS Collaboration}",
    title          = "{The ATLAS Experiment at the CERN Large Hadron Collider}",
    journal        = "JINST",
    volume         = "3",
    year           = "2008",
    pages          = "S08003",
    doi            = "10.1088/1748-0221/3/08/S08003",
    primaryClass   = "hep-ex",
}

@Article{SOFT-2010-01,
    author         = "{ATLAS Collaboration}",
    title          = "{The ATLAS Simulation Infrastructure}",
    journal        = "Eur. Phys. J. C",
    volume         = "70",
    year           = "2010",
    pages          = "823",
    doi            = "10.1140/epjc/s10052-010-1429-9",
    eprint         = "1005.4568",
    archivePrefix  = "arXiv",
    primaryClass   = "physics.ins-det",
}

@Article{PERF-2012-01,
    author         = "{ATLAS Collaboration}",
    title          = "{Jet energy measurement and its systematic uncertainty in proton--proton collisions at \(\sqrt{s} = 7\,\text{TeV}\) with the ATLAS detector}",
    journal        = "Eur. Phys. J. C",
    volume         = "75",
    year           = "2015",
    pages          = "17",
    doi            = "10.1140/epjc/s10052-014-3190-y",
    reportNumber   = "CERN-PH-EP-2013-222",
    eprint         = "1406.0076",
    archivePrefix  = "arXiv",
    primaryClass   = "hep-ex",
}

@Article{STDM-2012-19,
    author         = "{ATLAS Collaboration}",
    title          = "{Measurement of dijet azimuthal decorrelations in \(pp\) collisions at \(\sqrt{s} = 8\,\text{TeV}\) with the ATLAS detector and determination of the strong coupling}",
    journal        = "Phys. Rev. D",
    volume         = "98",
    year           = "2018",
    pages          = "092004",
    doi            = "10.1103/PhysRevD.98.092004",
    reportNumber   = "CERN-EP-2017-282",
    eprint         = "1805.04691",
    archivePrefix  = "arXiv",
    primaryClass   = "hep-ex",
}

@Article{STDM-2014-03,
    author         = "{ATLAS Collaboration}",
    title          = "{Measurement of transverse energy--energy correlations in multi-jet events in \(pp\) collisions at \(\sqrt{s} = 7\,\text{TeV}\) using the ATLAS detector and determination of the strong coupling constant \(\alpha_{\text{s}}(m_Z)\)}",
    journal        = "Phys. Lett. B",
    volume         = "750",
    year           = "2015",
    pages          = "427",
    doi            = "10.1016/j.physletb.2015.09.050",
    reportNumber   = "CERN-PH-EP-2015-177",
    eprint         = "1508.01579",
    archivePrefix  = "arXiv",
    primaryClass   = "hep-ex",
}

@Article{STDM-2016-10,
    author         = "{ATLAS Collaboration}",
    title          = "{Determination of the strong coupling constant \(\alpha_{\text{s}}\) from transverse energy-energy correlations in multijet events at \(\sqrt{s} = 8\,\text{TeV}\) using the ATLAS detector}",
    journal        = "Eur. Phys. J. C",
    volume         = "77",
    year           = "2017",
    pages          = "872",
    doi            = "10.1140/epjc/s10052-017-5442-0",
    reportNumber   = "CERN-EP-2017-093",
    eprint         = "1707.02562",
    archivePrefix  = "arXiv",
    primaryClass   = "hep-ex",
}

@Article{TRIG-2016-01,
    author         = "{ATLAS Collaboration}",
    title          = "{Performance of the ATLAS trigger system in 2015}",
    journal        = "Eur. Phys. J. C",
    volume         = "77",
    year           = "2017",
    pages          = "317",
    doi            = "10.1140/epjc/s10052-017-4852-3",
    reportNumber   = "CERN-EP-2016-241",
    eprint         = "1611.09661",
    archivePrefix  = "arXiv",
    primaryClass   = "hep-ex",
}

@Article{DAPR-2018-01,
    author         = "{ATLAS Collaboration}",
    title          = "{ATLAS data quality operations and performance for 2015--2018 data-taking}",
    journal        = "JINST",
    volume         = "15",
    year           = "2020",
    pages          = "P04003",
    doi            = "10.1088/1748-0221/15/04/P04003",
    reportNumber   = "CERN-EP-2019-207",
    eprint         = "1911.04632",
    archivePrefix  = "arXiv",
    primaryClass   = "physics.ins-det",
}

@Article{PIX-2018-001,
    author         = "Abbott, B. and others",
    title          = "{Production and integration of the ATLAS Insertable B-Layer}",
    journal        = "JINST",
    volume         = "13",
    year           = "2018",
    pages          = "T05008",
    doi            = "10.1088/1748-0221/13/05/T05008",
    eprint         = "1803.00844",
    archivePrefix  = "arXiv",
    primaryClass   = "physics.ins-det",
}

@Booklet{ATL-SOFT-PUB-2021-001,
    author         = "{ATLAS Collaboration}",
    title          = "{The ATLAS Collaboration Software and Firmware}",
    howpublished   = "{ATL-SOFT-PUB-2021-001}",
    url            = "https://cds.cern.ch/record/2767187",
    year           = "2021",
}

@Booklet{ATL-SOFT-PUB-2021-003,
    author         = "{ATLAS Collaboration}",
    title          = "{ATLAS Computing Acknowledgements}",
    howpublished   = "{ATL-SOFT-PUB-2021-003}",
    url            = "https://cds.cern.ch/record/2776662",
    year           = "2021",
}

@Report{ATLAS-TDR-19,
    author         = "{ATLAS Collaboration}",
    title          = "{ATLAS Insertable B-Layer: Technical Design Report}",
    type           = "ATLAS-TDR-19; CERN-LHCC-2010-013",
    year           = "2010",
    url            = "https://cds.cern.ch/record/1291633",
    related        = "ATLAS-TDR-19-addm",
    relatedstring  = "Addendum:",
}

@Article{CMS-QCD-11-003,
    author         = "{CMS Collaboration}",
    title          = "{Measurement of the ratio of the inclusive 3-jet cross section to the inclusive 2-jet cross section in \(pp\) collisions at \(\sqrt{s} = 7\,\text{TeV}\) and first determination of the strong coupling constant in the TeV range}",
    journal        = "Eur. Phys. J. C",
    volume         = "73",
    year           = "2013",
    pages          = "2604",
    doi            = "10.1140/epjc/s10052-013-2604-6",
    reportNumber   = "CERN-PH-EP-2013-057",
    eprint         = "1304.7498",
    archivePrefix  = "arXiv",
    primaryClass   = "hep-ex",
}

@Article{CMS-SMP-12-027,
    author         = "{CMS Collaboration}",
    title          = "{Measurement of the inclusive 3-jet production differential cross section in proton--proton collisions at \(7\,\text{TeV}\) and determination of the strong coupling constant in the TeV\,range}",
    journal        = "Eur. Phys. J. C",
    volume         = "75",
    year           = "2015",
    pages          = "186",
    doi            = "10.1140/epjc/s10052-015-3376-y",
    reportNumber   = "CERN-PH-EP-2014-287",
    eprint         = "1412.1633",
    archivePrefix  = "arXiv",
    primaryClass   = "hep-ex",
}

@Article{CMS-SMP-14-001,
    author         = "{CMS Collaboration}",
    title          = "{Measurement and QCD analysis of double-differential inclusive jet cross sections in \(pp\) collisions at \(\sqrt{s} = 8\,\text{TeV}\) and cross section ratios to \(2.76\) and \(7\,\text{TeV}\)}",
    journal        = "JHEP",
    volume         = "03",
    year           = "2017",
    pages          = "156",
    doi            = "10.1007/JHEP03(2017)156",
    reportNumber   = "CERN-EP-2016-196",
    eprint         = "1609.05331",
    archivePrefix  = "arXiv",
    primaryClass   = "hep-ex",
}

@Article{CMS-TOP-18-004,
    author         = "{CMS Collaboration}",
    title          = "{Measurement of \(t\bar{t}\) normalised multi-differential cross sections in \(pp\) collisions at \(\sqrt{s} = 13\,\text{TeV}\), and simultaneous determination of the strong coupling strength, top quark pole mass, and parton distribution functions}",
    journal        = "Eur. Phys. J. C",
    volume         = "80",
    year           = "2020",
    pages          = "658",
    doi            = "10.1140/epjc/s10052-020-7917-7",
    reportNumber   = "CERN-EP-2019-028",
    eprint         = "1904.05237",
    archivePrefix  = "arXiv",
    primaryClass   = "hep-ex",
}

@Booklet{ATLAS-CONF-2015-029,
    author         = "{ATLAS Collaboration}",
    title          = "{Selection of jets produced in \(13~\text{TeV}\) proton--proton collisions with the ATLAS detector}",
    howpublished   = "{ATLAS-CONF-2015-029}",
    url            = "https://cds.cern.ch/record/2037702",
    year           = "2015",
}

@Booklet{ATL-PHYS-PUB-2014-021,
    author         = "{ATLAS Collaboration}",
    title          = "{ATLAS Pythia~8 tunes to \(7~\text{TeV}\) data}",
    howpublished   = "{ATL-PHYS-PUB-2014-021}",
    url            = "https://cds.cern.ch/record/1966419",
    year           = "2014",
}

@Booklet{ATL-PHYS-PUB-2016-017,
    author         = "{ATLAS Collaboration}",
    title          = "{The Pythia~8 A3 tune description of ATLAS minimum bias and inelastic measurements incorporating the Donnachie--Landshoff diffractive model}",
    howpublished   = "{ATL-PHYS-PUB-2016-017}",
    url            = "https://cds.cern.ch/record/2206965",
    year           = "2016",
}

@Article{EEC-TH-00,
    author         = "{C. L. Basham, L. S. Brown, S. D. Ellis, S. T. Love}",
    title          = "{Energy Correlations in Electron-Positron Annihilation: Testing Quantum Chromodynamics}",
    journal        = "Phys. Rev. Lett.",
    volume         = "41",
    year           = "1978",
    pages          = "1585",
    doi            = "10.1103/PhysRevLett.41.1585",
}

@Article{EEC-TH-01,
    author         = "{C. L. Basham, L. S. Brown, S. D. Ellis, S. T. Love}",
    title          = "{Energy correlations in electron-positron annihilation in quantum chromodynamics: Asymptotically free perturbation theory}",
    journal        = "Phys. Rev. D",
    volume         = "19",
    year           = "1979",
    pages          = "2018",
    doi            = "10.1103/PhysRevD.19.2018",
}

@Article{EEC-TH-02,
    author         = "{A. Ali, F. Barreiro}",
    title          = "{An $\mathcal{O}(\alpha_{\mathrm{s}})^{2}$ calculation of energy-energy correlation in $e^{+} e^{-}$ annihilation and comparison with experimental data}",
    journal        = "Phys. Lett. B",
    volume         = "118",
    year           = "1982",
    pages          = "155",
    doi            = "10.1016/0370-2693(82)90621-9",
}

@Article{EEC-TH-03,
    author         = "{A. Ali, F. Barreiro}",
    title          = "{Energy-energy correlations in $e^{+} e^{-}$ annihilation}",
    journal        = "Nucl. Phys. B",
    volume         = "236",
    year           = "1984",
    pages          = "269",
    doi            = "10.1016/0550-3213(84)90536-4",
}

@Article{EEC-TH-04,
    author         = "{D. G. Richards, W. J. Stirling, S.D. Ellis}",
    title          = "{Second order corrections to the energy-energy correlation function in quantum chromodynamics}",
    journal        = "Phys. Lett. B",
    volume         = "119",
    year           = "1982",
    pages          = "193",
    doi            = "10.1016/0370-2693(82)90275-1",
}

@Article{EEC-TH-04b,
    author         = "{E.W.N. Glover, M. R. Sutton}",
    title          = "{The energy-energy correlation function revisited}",
    journal        = "Phys. Lett. B",
    volume         = "342",
    year           = "1995",
    pages          = "375",
    doi            = "10.1016/0370-2693(94)01354-F",
    eprint         = "hep-ph/9410234",
    archivePrefix  = "arXiv",
}

@Article{EEC-TH-05,
    author         = "{S. D. Ellis, D. G. Richards, W. J. Stirling}",
    title          = "{Fixed order perturbation theory and leading logarithms in energy-energy correlations}",
    journal        = "Phys. Lett. B",
    volume         = "136",
    year           = "1984",
    pages          = "99",
    doi            = "10.1016/0370-2693(84)92064-1",
}

@Article{CERN-TH-3800,
    author         = "{A. Ali, E. Pietarinen, W. J. Stirling}",
    title          = "{Transverse energy-energy correlations: A test of perturbative QCD for the proton-antiproton collider}",
    journal        = "Phys. Lett. B",
    volume         = "141",
    year           = "1984",
    pages          = "447",
    doi            = "10.1016/0370-2693(84)90283-1",
    reportNumber   = "CERN-TH-3800",
}

@Article{TEEC-TH-00,
    author         = "{G. Altarelli}",
    title          = "{The Development of Perturbative QCD}",
    journal        = "World Scientific",
    year           = "1994",
    doi            = "10.1142/2300",
}

@Article{TEEC-TH-01,
    author         = "{A. Ali, F. Barreiro, J. Llorente, W. Wang}",
    title          = "{Transverse energy-energy correlations in next-to-leading order in $\alpha_{\mathrm{s}}$ at the LHC}",
    journal        = "Phys. Rev. D",
    volume         = "86",
    year           = "2012",
    pages          = "114017",
    doi            = "10.1103/PhysRevD.86.114017",
    eprint         = "1205.1689",
    archivePrefix  = "arXiv",
    primaryClass   = "hep-ph",
}

@Article{TEEC-TH-02,
    author         = "{A. Gao, H. T. Li, I. Moult, H. X. Zhu}",
    title          = "{Precision QCD Event Shapes at Hadron Colliders: The Transverse Energy-Energy Correlator in the Back-to-Back Limit}",
    journal        = "Phys. Rev. Lett.",
    volume         = "123",
    year           = "2019",
    pages          = "062001",
    doi            = "10.1103/PhysRevLett.123.062001",
    eprint         = "1901.04497",
    archivePrefix  = "arXiv",
    primaryClass   = "hep-ph",
}

@Article{IPPP/01/48,
    author         = "{Z. Nagy}",
    title          = "{Three-Jet Cross Sections in Hadron-Hadron Collisions at Next-to-Leading Order}",
    journal        = "Phys. Rev. Lett.",
    volume         = "88",
    year           = "2002",
    pages          = "122003",
    doi            = "10.1103/PhysRevLett.88.122003",
    reportNumber   = "IPPP/01/48, DCPT/01/96",
    eprint         = "hep-ph/0110315",
    archivePrefix  = "arXiv",
}

@Article{TEEC-TH-03,
    author         = "{Z. Nagy}",
    title          = "{Next-to-leading order calculation of three-jet observables in hadron-hadron collisions}",
    journal        = "Phys. Rev. D",
    volume         = "68",
    year           = "2003",
    pages          = "094002",
    doi            = "10.1103/PhysRevD.68.094002",
    eprint         = "hep-ph/0307268",
    archivePrefix  = "arXiv",
}

@Article{TEEC-TH-04,
    author         = "{D. E. Kaplan, M. D. Schwartz}",
    title          = "{Constraining Light Colored Particles with Event Shapes}",
    journal        = "Phys. Rev. Lett.",
    volume         = "101",
    year           = "2008",
    pages          = "022002",
    doi            = "10.1103/PhysRevLett.101.022002",
    eprint         = "0804.2477",
    archivePrefix  = "arXiv",
    primaryClass   = "hep-ph",
}

@Article{TEEC-TH-05,
    author         = "{D. Becciolini, M. Gillioz, M. Nardecchia, F. Sannino, M. Spannowsky}",
    title          = "{Constraining new colored matter from the ratio of 3 to 2 jets cross sections at the LHC}",
    journal        = "Phys. Rev. D",
    volume         = "91",
    year           = "2015",
    pages          = "015010",
    doi            = "10.1103/PhysRevD.91.015010",
    eprint         = "1403.7411",
    archivePrefix  = "arXiv",
    primaryClass   = "hep-ph",
}

@Article{TEEC-TH-06,
    author         = "{J. Llorente, B. Nachman}",
    title          = "{Limits on new coloured fermions using precision jet data from the Large Hadron Collider}",
    journal        = "Nucl. Phys. B",
    volume         = "936",
    year           = "2018",
    pages          = "106",
    doi            = "10.1016/j.nuclphysb.2018.09.008",
    eprint         = "1807.00894",
    archivePrefix  = "arXiv",
    primaryClass   = "hep-ph",
}

@Article{Matchbox,
    author         = "{S. Pl\"{a}tzer, S. Gieseke}",
    title          = "{Dipole Showers and Automated NLO Matching in Herwig++}",
    journal        = "Eur. Phys. J. C",
    volume         = "72",
    year           = "2012",
    pages          = "2187",
    doi            = "10.1140/epjc/s10052-012-2187-7",
    eprint         = "1109.6256",
    archivePrefix  = "arXiv",
    primaryClass   = "hep-ph",
}

@Article{GEANT4,
    author         = "{S. Agostinelli \emph{et al.}}",
    title          = "{GEANT4 - a simulation toolkit}",
    journal        = "Nucl. Instrum. Meth. A",
    volume         = "506",
    year           = "2003",
    pages          = "250",
    doi            = "10.1016/S0168-9002(03)01368-8",
}

@Article{MMHT2014,
    author         = "{L. A. Harland-Lang, A. D. Martin, P. Motylinski, R. S. Thorne}",
    title          = "{Parton distributions in the LHC era: MMHT 2014 PDFs}",
    journal        = "Eur. Phys. J. C",
    volume         = "75",
    year           = "2015",
    pages          = "204",
    doi            = "10.1140/epjc/s10052-015-3397-6",
    eprint         = "1412.3989",
    archivePrefix  = "arXiv",
    primaryClass   = "hep-ph",
}

@Article{LHAPDF,
    author         = "{A. Buckley \emph{et al.}}",
    title          = "{LHAPDF6: parton density access in the LHC precision era}",
    journal        = "Eur. Phys. J. C",
    volume         = "75",
    year           = "2015",
    pages          = "132",
    doi            = "10.1140/epjc/s10052-015-3318-8",
    eprint         = "1412.7420",
    archivePrefix  = "arXiv",
    primaryClass   = "hep-ph",
}

@Article{Tune4C,
    author         = "{R. Corke and T. Sjöstrand}",
    title          = "{Interleaved parton showers and tuning prospects}",
    journal        = "JHEP",
    volume         = "03",
    year           = "2011",
    pages          = "032",
    doi            = "10.1007/JHEP03(2011)032",
    archivePrefix = "arXiv",
    eprint = "1011.1759",
    primaryClass = "hep-ph"
}

@Article{herwigTunes,
  author = "{S. Gieseke, C. R\"ohr and A. Si\'odmok}",
  title = "{Colour reconnections in Herwig++}",
  journal = "Eur. Phys. J. C",
  volume = "72",
  pages = "2225",
  year = "2012",
  archivePrefix = "arXiv",
  primaryClass = "hep-ph",
  eprint = "1206.0041",
  doi = "10.1140/epjc/s10052-012-2225-5"
}

@Article{CERN-TH-2018-159,
    author         = "{J. Currie \emph{et al.}}",
    title          = "{Infrared sensitivity of single jet inclusive production at hadron colliders}",
    journal        = "JHEP",
    volume         = "10",
    year           = "2018",
    pages          = "155",
    doi            = "10.1007/JHEP10(2018)155",
    reportNumber   = "IPPP/18/38, ZU-TH 24/18, CFTP/18-010, CERN-TH-2018-159",
    eprint         = "1807.03692",
    archivePrefix  = "arXiv",
    primaryClass   = "hep-ph",
}

@article{PDG2022,
    author = "{Particle Data Group}",
    title = "{Review of Particle Physics}",
    doi = "10.1093/ptep/ptac097",
    journal = "PTEP",
    volume = "2022",
    pages = "083C01",
    year = "2022"
}

@Article{D0-1,
    author         = "{D0 Collaboration}",
    title          = "{Measurement of angular correlations of jets at $\sqrt{s}=1.96\;\mbox{TeV}$ and determination of the strong coupling at high momentum transfers}",
    journal        = "Phys. Rev. B",
    volume         = "718",
    year           = "2012",
    pages          = "56",
    doi            = "10.1016/j.physletb.2012.10.003",
    eprint         = "1207.4957",
    archivePrefix  = "arXiv",
    primaryClass   = "hep-ex",
}

@Article{D0-2,
    author         = "{D0 Collaboration}",
    title          = "{Determination of the strong coupling constant from the inclusive jet cross section in $p\bar{p}$ collisions at $\sqrt{s}=1.96\;\mbox{TeV}$}",
    journal        = "Phys. Rev. D",
    volume         = "80",
    year           = "2009",
    pages          = "111107",
    doi            = "10.1103/PhysRevD.80.111107",
    eprint         = "0911.2710",
    archivePrefix  = "arXiv",
    primaryClass   = "hep-ex",
}

@Article{PLUTO,
   author="{PLUTO Collaboration}",
   journal="Phys. Lett. B",
   volume="99",
   pages="292",
   year=1981,
   doi="10.1016/0370-2693(81)91128-X",
   title="{Energy-energy correlations in $e^+e^-$ annihilation into hadrons}"
}

@Article{MARKII,
   author="{MARKII Collaboration}",
   journal="Phys. Rev. Lett.",
   volume="49",
   pages="521",
   year=1982,
   doi="10.1103/PhysRevLett.49.521",
   title="{Measurement of Energy Correlations in $e^+e^- \to $ Hadrons}"
}

@Article{MARKJ,
   author="{MARKJ Collaboration}",
   journal="Phys. Rev. Lett.",
   volume="50",
   pages="2051",
   year=1983,
   doi="10.1103/PhysRevLett.50.2051",
   title="{Model-Independent Second-Order Determination of the Strong-Coupling Constant $\alpha_s$}"
}

@Article{CELLO,
   author="{CELLO Collaboration}",
   journal="Phys. Lett. B",
   volume="138",
   pages="311",
   year=1984,
   doi="10.1016/0370-2693(84)91667-8",
   title="{On the model dependence of the determination of the strong coupling constant in second order QCD from $e^+e^-$ annihilation into hadrons}"
}

@Article{JADE,
   author="{JADE Collaboration}",
   journal="Z. Phys. C",
   volume="25",
   pages="231",
   year=1984,
   doi="10.1007/BF01547922",
   title="{Measurements of energy correlations in $e^+e^-\to$ hadrons}"
}

@Article{MAC,
   author="{MAC Collaboration}",
   journal="Phys. Rev. D",
   volume="31",
   pages="2724",
   year=1985,
   doi="10.1103/PhysRevD.31.2724",
   title="{Measurement of energy-energy correlations in $e^+e^-\to$ hadrons at $\sqrt{s} = 29 \GeV$}"
}

@Article{TASSO,
   author="{TASSO Collaboration}",
   journal= "Z. Phys. C",
   volume="36",
   pages="349",
   year=1987,
   doi="10.1007/BF01573928",
   title="{A study of energy-energy correlations between 12 and 46.8 GeV c.m. energies}"
}

@Article{TOPAZ,
   author="{TOPAZ Collaboration}",
   journal="Phys. Lett. B",
   volume="227",
   pages="495",
   year=1989,
   doi="10.1016/0370-2693(89)90969-6",
   title="{Measurements of $\alpha_s$ in $e^+e^-$ annihilation at $\sqrt{s} =53.3 \GeV$ and $59.5 \GeV$}"
}

@Article{DELPHI,
   author="{DELPHI Collaboration}",
   journal="Phys. Lett. B",
   volume="252",
   pages="149",
   year=1990,
   doi="10.1016/0370-2693(90)91097-U",
   title="{Energy-energy correlations in hadronic final states from $\Zboson^0$ decays}"
}

@Article{OPAL,
   author="{OPAL Collaboration}",
   journal="Phys. Lett. B",
   volume="252",
   pages="159",
   year=1990,
   doi="10.1016/0370-2693(90)91098-V",
   title="{A measurement of energy correlations and a determination of $\alpha_s(M^2_{\Zboson^0})$ in $e^+e^-$ annihilations at $\sqrt{s} =91 \GeV$}"
}

@Article{ALEPH,
   author="{ALEPH Collaboration}",
   journal="Phys. Lett. B",
   volume = "257",
   pages="479",
   year=1991,
   doi="10.1016/0370-2693(91)91926-M",
   title="{Measurement of $\alpha_s$ from the structure of particle clusters produced in hadronic $\Zboson$ decays}"
}

@Article{L3,
   author="{L3 Collaboration}",
   journal="Phys. Lett. B",
   volume = "257",
   pages = "469",
   year=1991,
   doi="10.1016/0370-2693(91)91925-L",
   title="{Determination of $\alpha_s$ from energy-energy correlations measured on the $\Zboson^0$ resonance}"
}

@Article{SLD,
   author="{SLD Collaboration}",
   journal="Phys. Rev. D",
   volume="51",
   pages = "962",
   year=1995,
   doi="10.1103/PhysRevD.51.962",
   title="{Measurement of $\alpha_s(M^2_\Zboson)$ from hadronic event observables at the $\Zboson^0$ resonance}"
}

@Article{eecSLD,
   author = "{SLD Collaboration}",
   journal = "Phys. Rev. D",
   volume = "50",
   pages = "5580",
   year = "1994",
   doi="10.1103/PhysRevD.50.5580",
   title="Measurement of $\alpha_s$ from energy-energy correlations at the $Z^0$ resonance"
}

\clearpage
 
 
\begin{flushleft}
\hypersetup{urlcolor=black}
{\Large The ATLAS Collaboration}

\bigskip

\AtlasOrcid[0000-0002-6665-4934]{G.~Aad}$^\textrm{\scriptsize 102}$,
\AtlasOrcid[0000-0002-5888-2734]{B.~Abbott}$^\textrm{\scriptsize 120}$,
\AtlasOrcid[0000-0002-1002-1652]{K.~Abeling}$^\textrm{\scriptsize 55}$,
\AtlasOrcid[0000-0002-8496-9294]{S.H.~Abidi}$^\textrm{\scriptsize 29}$,
\AtlasOrcid[0000-0002-9987-2292]{A.~Aboulhorma}$^\textrm{\scriptsize 35e}$,
\AtlasOrcid[0000-0001-5329-6640]{H.~Abramowicz}$^\textrm{\scriptsize 151}$,
\AtlasOrcid[0000-0002-1599-2896]{H.~Abreu}$^\textrm{\scriptsize 150}$,
\AtlasOrcid[0000-0003-0403-3697]{Y.~Abulaiti}$^\textrm{\scriptsize 117}$,
\AtlasOrcid[0000-0003-0762-7204]{A.C.~Abusleme~Hoffman}$^\textrm{\scriptsize 137a}$,
\AtlasOrcid[0000-0002-8588-9157]{B.S.~Acharya}$^\textrm{\scriptsize 69a,69b,p}$
% ...
\bigskip
\\

% ...
$^{29}$Physics Department, Brookhaven National Laboratory, Upton NY; United States of America.\\
% ...
$^{35}$$^{(a)}$Facult\'e des Sciences Ain Chock, R\'eseau Universitaire de Physique des Hautes Energies - Universit\'e Hassan II, Casablanca;$^{(b)}$Facult\'{e} des Sciences, Universit\'{e} Ibn-Tofail, K\'{e}nitra;$^{(c)}$Facult\'e des Sciences Semlalia, Universit\'e Cadi Ayyad, LPHEA-Marrakech;$^{(d)}$LPMR, Facult\'e des Sciences, Universit\'e Mohamed Premier, Oujda;$^{(e)}$Facult\'e des sciences, Universit\'e Mohammed V, Rabat;$^{(f)}$Institute of Applied Physics, Mohammed VI Polytechnic University, Ben Guerir; Morocco.\\
% ...
$^{55}$II. Physikalisches Institut, Georg-August-Universit\"{a}t G\"ottingen, G\"ottingen; Germany.\\
% ...
$^{69}$$^{(a)}$INFN Gruppo Collegato di Udine, Sezione di Trieste, Udine;$^{(b)}$ICTP, Trieste;$^{(c)}$Dipartimento Politecnico di Ingegneria e Architettura, Universit\`a di Udine, Udine; Italy.\\
% ...
$^{102}$CPPM, Aix-Marseille Universit\'e, CNRS/IN2P3, Marseille; France.\\
% ...
$^{117}$Department of Physics, New York University, New York NY; United States of America.\\
% ...
$^{120}$Homer L. Dodge Department of Physics and Astronomy, University of Oklahoma, Norman OK; United States of America.\\
% ...
$^{137}$$^{(a)}$Departamento de F\'isica, Pontificia Universidad Cat\'olica de Chile, Santiago;$^{(b)}$Millennium Institute for Subatomic physics at high energy frontier (SAPHIR), Santiago;$^{(c)}$Instituto de Investigaci\'on Multidisciplinario en Ciencia y Tecnolog\'ia, y Departamento de F\'isica, Universidad de La Serena;$^{(d)}$Universidad Andres Bello, Department of Physics, Santiago;$^{(e)}$Instituto de Alta Investigaci\'on, Universidad de Tarapac\'a, Arica;$^{(f)}$Departamento de F\'isica, Universidad T\'ecnica Federico Santa Mar\'ia, Valpara\'iso; Chile.\\
% ...
$^{150}$Department of Physics, Technion, Israel Institute of Technology, Haifa; Israel.\\
$^{151}$Raymond and Beverly Sackler School of Physics and Astronomy, Tel Aviv University, Tel Aviv; Israel.\\
% ...
$^{a}$ Also Affiliated with an institute covered by a cooperation agreement with CERN.\\
$^{b}$ Also at An-Najah National University, Nablus; Palestine.\\
$^{c}$ Also at Borough of Manhattan Community College, City University of New York, New York NY; United States of America.\\
$^{d}$ Also at Bruno Kessler Foundation, Trento; Italy.\\
$^{e}$ Also at Center for High Energy Physics, Peking University; China.\\
$^{f}$ Also at Center for Interdisciplinary Research and Innovation (CIRI-AUTH), Thessaloniki ; Greece.\\
% ...
$^{p}$ Also at Department of Physics, King's College London, London; United Kingdom.\\
% ...

\end{flushleft}


\end{document}